\documentclass[%
 reprint,
superscriptaddress,
 amsmath,amssymb,
 aps,
floatfix,
]{revtex4-2}

\usepackage{graphicx}
\usepackage{dcolumn}
\usepackage{bm}
\usepackage{amsmath,amssymb}
\usepackage{textcomp,mathcomp}
\usepackage{color}
\usepackage{url}
\begin{document}

\preprint{APS/123-QED}

\title{Coupled structural and electronic evolution under pressure in CuIr$_2$Se$_4$, CuRh$_2$S$_4$, and CuRh$_2$Se$_4$}
\author{M. Emi}
\affiliation{Department of Applied Physics, Nagoya University, Nagoya 464-8603, Japan}
\affiliation{Graduate School of Environmental, Life, Natural Science and Technology, Okayama University, Okayama 700-8530, Japan}

\author{M. Shiomi}
\affiliation{Department of Applied Physics, Nagoya University, Nagoya 464-8603, Japan}
\author{K. Kojima}

\affiliation{Department of Applied Physics, Nagoya University, Nagoya 464-8603, Japan}
\affiliation{Graduate School of Environmental, Life, Natural Science and Technology, Okayama University, Okayama 700-8530, Japan}
\affiliation{The Institute for Solid State Physics, The University of Tokyo, Kashiwa 277-8581, Japan}

\author{K. Sugimoto}
\affiliation{Department of Physics, Kitasato University, Minami-ku, Sagamihara, Kanagawa 252-0373, Japan}

\author{T. Karasawa}
\affiliation{Department of Engineering Science, The University of Electro-Communications, Chofu, Tokyo 182-8585, Japan}

\author{H. Suzuki}
\affiliation{Department of Engineering Science, The University of Electro-Communications, Chofu, Tokyo 182-8585, Japan}
\author{M. Takahashi}
\affiliation{Department of Engineering Science, The University of Electro-Communications, Chofu, Tokyo 182-8585, Japan}
\author{K. Oka}
\affiliation{Japan Synchrotron Radiation Research Institute (JASRI), SPring-8, Hyogo 679-5198, Japan}

\author{H. Kadobayashi}
\affiliation{Japan Synchrotron Radiation Research Institute (JASRI), SPring-8, Hyogo 679-5198, Japan}
\author{S. Kawaguchi-Imada}
\affiliation{Office of Institutional Advancement and Communications, Kyoto University, Yoshidahonmachi, Sakyo-ku, Kyoto 606-8501, Japan}
\affiliation{Japan Synchrotron Radiation Research Institute (JASRI), SPring-8, Hyogo 679-5198, Japan}

\author{N. Hirao}
\affiliation{Japan Synchrotron Radiation Research Institute (JASRI), SPring-8, Hyogo 679-5198, Japan}
\author{T. Ohashi}
\affiliation{Department of Applied Physics, Nagoya University, Nagoya 464-8603, Japan}
\affiliation{Graduate School of Environmental, Life, Natural Science and Technology, Okayama University, Okayama 700-8530, Japan}

\author{D. Ito}
\affiliation{Department of Applied Physics, Nagoya University, Nagoya 464-8603, Japan}
\affiliation{Graduate School of Environmental, Life, Natural Science and Technology, Okayama University, Okayama 700-8530, Japan}

\author{T. Kubo}
\affiliation{Department of Applied Physics, Nagoya University, Nagoya 464-8603, Japan}
\affiliation{Graduate School of Environmental, Life, Natural Science and Technology, Okayama University, Okayama 700-8530, Japan}

\author{M. Matsushita}
\affiliation{Department of Quantum Matter, AdSE, Hiroshima University, Higashihiroshima 739-8530, Japan}

\author{M. Nohara}
\affiliation{Department of Quantum Matter, AdSE, Hiroshima University, Higashihiroshima 739-8530, Japan}

\author{K. Matsubayashi}\thanks{Corresponding author}\email{k.matsubayashi@uec.ac.jp}
\affiliation{Department of Engineering Science, The University of Electro-Communications, Chofu, Tokyo 182-8585, Japan}

\author{N. Katayama}\thanks{Corresponding author}\email{knaoyuki@okayama-u.ac.jp}
\affiliation{Department of Applied Physics, Nagoya University, Nagoya 464-8603, Japan}
\affiliation{Graduate School of Environmental, Life, Natural Science and Technology, Okayama University, Okayama 700-8530, Japan}

%
\date{\today}

\begin{abstract}
Spinel chalcogenides provide a platform for investigating the interplay among metallic, superconducting, and pressure-induced insulating states. Here, we combine synchrotron powder X-ray diffraction and electrical-resistivity measurements to investigate the pressure evolution of CuIr$_2$Se$_4$, CuRh$_2$S$_4$, and CuRh$_2$Se$_4$ over pressure ranges extending beyond those previously explored. High-pressure diffraction reveals closely related monoclinic supercells in all three compounds. For CuIr$_2$Se$_4$ and CuRh$_2$S$_4$, constrained profile fits based on structural models relaxed using density functional theory are compatible with Phase-IV-type bond-disproportionated structures, whereas the data for CuRh$_2$Se$_4$ establish a compatible monoclinic unit cell without resolving its atomic-scale ordering pattern. Insulating-like transport develops abruptly over a narrow pressure range in CuIr$_2$Se$_4$ but more gradually over broader pressure ranges in the Rh-based compounds, in close correspondence with their respective structural transformations. We also establish previously unreported bulk superconductivity in CuIr$_2$Se$_4$ at ambient pressure: zero resistance is attained at 0.29~K, and the accompanying ac diamagnetic response is consistent with nearly complete superconducting shielding. These results establish a close relationship between the formation of the high-pressure monoclinic phases and the evolution toward insulating transport, and demonstrate that transition-metal and chalcogen substitutions tune the characteristic pressure scales and the competition with superconductivity within a closely related structural framework.

\end{abstract}

\maketitle

\section{Introduction}

In spinel compounds with the general formula $AB_2X_4$, the transition-metal ions occupying the $B$ sites form a pyrochlore lattice, where geometrical frustration and interactions among spin, charge, orbital, and lattice degrees of freedom give rise to a wide variety of electronic states \cite{frustrate,spinels_review,spinels_review2}. The resulting competition among these states is particularly pronounced in mixed-valence spinels because multiple charge- and bond-ordering patterns can satisfy similar local constraints and can therefore be close in energy \cite{anderson}. Pressure is a particularly useful means of controlling the relative stability of these competing states because it continuously tunes the electronic and lattice structures without introducing chemical disorder \cite{pressure_without_disorder}.

LiV$_2$O$_4$ is a representative example. At ambient pressure, it exhibits heavy-fermion behavior, which is unusual for a $3d$-electron system \cite{LiV2O4_heavy1,LiV2O4_heavy2,LiV2O4_heavy3,LiV2O4_heavy4}. Under hydrostatic pressure, charge ordering is thought to emerge \cite{LiV2O4_hp_XRD_1,LiV2O4_hp_opt,LiV2O4_hp_NMR,LiV2O4_hp_EXAFS,LiV2O4_hp_XRD_2}. By contrast, uniaxial distortion induced by epitaxial strain stabilizes a different charge-ordering pattern \cite{LiV2O4_uniaxis}. This observation suggests that multiple charge-ordered states are energetically competitive on the pyrochlore lattice. It has been argued that fluctuations among these competing charge-ordered states may be responsible for the emergence of the heavy-fermion state at ambient pressure.

CuIr$_2$S$_4$ provides another representative example in which pressure selects among competing electronic and structural states. At ambient pressure and low temperatures, CuIr$_2$S$_4$ undergoes a metal--insulator transition into a bond-disproportionated state with Ir--Ir dimers, conventionally described within an idealized ionic picture in terms of nominal Ir$^{3+}$/Ir$^{4+}$ order \cite{CuIr2S4_CuIr2Se4_MIT,CuIr2S4_con1,CuIr2S4_radaelli,Takubo2005,ohashi}. Under high pressure, CuIr$_2$S$_4$ exhibits a series of structurally distinct phases, labeled Phases I--VII, as well as two distinct superconducting phases \cite{CuIr2S4_superconductivity}. Recent structural analysis has shown that Phase II has the same crystal structure as the ambient-pressure low-temperature phase, that Phase III corresponds to a coexistence region of Phases II and IV, and that Phase IV exhibits an Ir--Ir short-bond topology distinct from that of Phase II \cite{CuIr2S4_emi}. Within a nominal-valence representation, the Phase-IV topology is consistent with an Anderson-type local tetrahedron rule. These results indicate that multiple electronic and structural states are energetically competitive in CuIr$_2$S$_4$ and that different states are selected upon the application of pressure.

Related pressure-induced behavior has been reported in CuIr$_2$Se$_4$, CuRh$_2$S$_4$, and CuRh$_2$Se$_4$. CuRh$_2$S$_4$ and CuRh$_2$Se$_4$ are superconductors at ambient pressure \cite{CuRh2S4_SC1,CuRh2Se4_SC1,CuRh2S4_CuRh2Se4_SC1,CuRh2S4_CuRh2Se4_SC2,CuRh2S4_CuRh2Se4_SC3,CuRh2S4_CuRh2Se4_Heat}, whereas no superconducting transition has previously been reported in CuIr$_2$Se$_4$ down to 0.4~K \cite{CuIr2S4_CuIr2Se4_MIT,CuIr2Se4_heat_capacity,CuIrPtSe4_superconductivity}. Although all three compounds are metallic at ambient pressure, they exhibit pressure-induced metal--insulator transitions \cite{CuIr2S4_CuIr2Se4_hp_resis_1,CuIr2S4_CuIr2Se4_hp_resis_2,CuRh2S4_hp_con,CuIr2Se4_hp_cond_1,CuIr2Se4_hp_cond_2,CuIr2Se4_hp_Opt,CuRh2S4_CuIr2Se4_hp_opt,CuRh2S4_hp_con_XRD,CuRh2S4_CuIr2S4_hp_con,CuRh2Se4_hp_con1}. This is unusual because compression generally enhances orbital overlap and electronic bandwidth, thereby favoring metallicity \cite{MIT_review,HP_review,miao2020}. The emergence of insulating behavior under pressure therefore points to the stabilization of an ordered state rather than a simple bandwidth-controlled response.

These compounds provide a useful platform for comparison because substitution of Ir by Rh and S by Se modifies the spatial extent of the transition-metal $d$ orbitals, the metal--metal distance, and the metal--chalcogen hybridization while preserving the spinel framework and the same nominal mixed valence. Nevertheless, the high-pressure ordered states of these compounds have not been established crystallographically, and the origin of their different pressure responses remains unclear.

In this study, we investigate the crystal structures and electrical transport properties of CuIr$_2$Se$_4$, CuRh$_2$S$_4$, and CuRh$_2$Se$_4$ under pressure using synchrotron powder X-ray diffraction and electrical resistivity measurements over pressure ranges extending beyond those previously explored. By applying the structural-analysis framework developed for CuIr$_2$S$_4$ \cite{CuIr2S4_emi}, we establish closely related monoclinic high-pressure supercells in all three compounds. For CuIr$_2$Se$_4$ and CuRh$_2$S$_4$, constrained profile fits using structural models optimized within density functional theory (DFT) support compatibility with Phase-IV-type bond-disproportionated structures. For CuRh$_2$Se$_4$, the available diffraction data establish a compatible monoclinic unit cell, although its atomic-scale ordering pattern remains unresolved. The resistivity measurements further reveal marked material-dependent differences in the pressure and temperature ranges over which insulating behavior emerges, despite the similarity of their high-pressure crystallographic frameworks. We also establish previously unreported bulk superconductivity in CuIr$_2$Se$_4$ at ambient pressure through the concurrence of zero resistance at 0.29~K and an ac diamagnetic response consistent with nearly complete superconducting shielding. These results clarify how the transition-metal and chalcogen species influence the competition among metallic, superconducting, and pressure-induced insulating states.

\section{Experimental Methods}

Powder samples were synthesized by a solid-state reaction method. Stoichiometric amounts of Cu (99.99\%), Rh (99.95\%) or Ir (99.95\%), and S (99.999\%) or Se (99.999\%), as appropriate for each composition, were mixed and sealed in evacuated quartz tubes. The mixtures were heated at 800~$^\circ$C for CuRh$_2$S$_4$ and CuRh$_2$Se$_4$ and at 850~$^\circ$C for CuIr$_2$Se$_4$, and held at the respective temperatures for 3 days. Phase purity was assessed by Rietveld analysis of ambient-pressure powder X-ray diffraction data collected at BL02B2 of SPring-8. No impurity phases were detected in either CuIr$_2$Se$_4$ or CuRh$_2$S$_4$ within the experimental detection limit, whereas CuRh$_2$Se$_4$ contained a minor RhSe$_2$ impurity phase at approximately 2.3 wt\% (5.0 mol\%). For high-pressure electrical resistivity measurements, portions of the resulting powders were consolidated into pellets by spark plasma sintering (SPS; Syntex Inc.) under a pressure of 100 MPa. The sintering temperatures were 800~$^\circ$C for CuRh$_2$S$_4$, 500~$^\circ$C for CuRh$_2$Se$_4$, and 700~$^\circ$C for CuIr$_2$Se$_4$, with holding times of 10, 2, and 10 min, respectively.

Single-crystal specimens of CuRh$_2$S$_4$ and CuRh$_2$Se$_4$ used for the single-crystal X-ray diffraction measurements were prepared separately from the powder samples used for the other measurements by a solid-state reaction. Cu powder (99.9\%), Rh powder (99.9\%), and S powder (99.99\%) or Se powder (99.9\%) were weighed in the stoichiometric molar ratio of 1:2:4. The mixtures were loaded into quartz tubes and sealed.  The mixtures were heated at 850~$^\circ$C and held at this temperature for 10 days.

Temperature-dependent powder X-ray diffraction measurements were performed at BL02B2 of SPring-8. N$_2$- and He-gas blowers were used above and below 100 K, respectively. Single-crystal X-ray diffraction measurements were performed at 100 K under ambient pressure at BL02B1 of SPring-8 using an X-ray energy of $E = 40$ keV, with the sample temperature maintained using an N$_2$-gas blower.
CrysAlisPro was used for indexing and intensity extraction \cite{CrysAlisPro_2014}, and JANA2006 was used for data averaging and structural refinement \cite{jana}. High-pressure powder X-ray diffraction measurements were performed at BL10XU of SPring-8 \cite{BL10XU}. The X-ray energy was \textit{E} = 30 keV. An imaging plate was used as the detector (R-AXIS IV++). The powder sample was loaded into a diamond-anvil cell (DAC) together with ruby as a pressure marker \cite{Ruby_1,Ruby_2}, and the sample chamber was filled with helium as the pressure-transmitting medium. The culet diameter of the diamond anvil was 0.5 mm. SUS304 gaskets were used for CuIr$_2$Se$_4$ and CuRh$_2$Se$_4$, and a rhenium gasket was used for CuRh$_2$S$_4$. The two-dimensional powder X-ray diffraction images were converted into one-dimensional diffraction patterns using IPAnalyzer and PDIndexer \cite{IPanalyzer}. The diffraction data were analyzed using RIETAN-FP \cite{Rietan2}, and the crystal structures were visualized using VESTA \cite{VESTA}. For CuIr$_2$Se$_4$ and CuRh$_2$S$_4$, the relative monoclinic phase fraction $w_m$ in the two-phase coexistence region was obtained from two-phase profile refinements using RIETAN-FP. For CuRh$_2$Se$_4$, quantitative two-phase refinement was not reliable because of severe preferred orientation and pronounced peak broadening and overlap. The relative cubic-phase content $w_c$ was therefore estimated from the integrated intensity of the cubic (111) reflection normalized to that in the single-phase cubic region, and the relative monoclinic content was defined as $w_m=1-w_c$. Accordingly, the resulting $w_m$ values for CuRh$_2$Se$_4$ are used as semiquantitative measures of the progress of the structural transformation rather than as absolute phase fractions.

Ambient-pressure electrical resistivity was measured down to 2~K using a Physical Property Measurement System (PPMS) and below 2~K using an Oxford Instruments Heliox $^3$He cryostat. The dc magnetic susceptibilities of CuRh$_2$S$_4$ and CuRh$_2$Se$_4$ were measured using a Magnetic Property Measurement System (MPMS). For CuIr$_2$Se$_4$, the ac magnetic-susceptibility signal was measured at a fixed frequency of 113~Hz, with the signal from the pickup coil detected using a lock-in amplifier. A Pb reference with a volume comparable to that of the CuIr$_2$Se$_4$ specimen was measured under identical conditions. The superconducting shielding fraction was estimated by comparing the changes in the ac susceptibility signals of CuIr$_2$Se$_4$ and Pb across their respective superconducting transitions.

High-pressure electrical resistivity measurements were carried out using an opposed-anvil cell \cite{Kitagawa}. Nonmagnetic tungsten carbide anvils with a culet diameter of 3.0~mm and a NiCrAl alloy gasket with a hole diameter of 1.8~mm were used. The pressure was determined from the superconducting transition temperature of Pb. Glycerin was used as the pressure-transmitting medium for CuIr$_2$Se$_4$, whereas argon was used for CuRh$_2$Se$_4$. For CuRh$_2$S$_4$, glycerin was used at pressures up to 2.0 GPa and argon at 2.9 GPa and above.

First-principles calculations based on DFT were performed using the Quantum ESPRESSO package \cite{calc_1,calc_2,calc_3,calc_4,calc_5}. The Perdew--Burke--Ernzerhof generalized-gradient approximation revised for solids (PBEsol) and Kresse--Joubert-type projector--augmented-wave pseudopotentials were employed. During the structural optimizations, the cubic symmetry of the ambient-pressure phase and the base-centered monoclinic symmetry of the high-pressure phase were preserved, while the lattice parameters and internal atomic coordinates were fully relaxed. The kinetic-energy cutoff for the plane-wave basis was set to 150~Ry. Brillouin-zone integrations were performed using $16 \times 16 \times 16$ and $8 \times 8 \times 8$ k-point meshes for the ambient- and high-pressure phases, respectively. The optimized structures of the ambient-pressure phase were subsequently used to calculate the density of states (DOS) and Fermi surfaces, with spin--orbit coupling included.

\section{Results and Discussion}
\subsection{High-pressure structural evolution of CuIr$_2$Se$_4$}

\begin{figure}
\includegraphics[width=85mm]{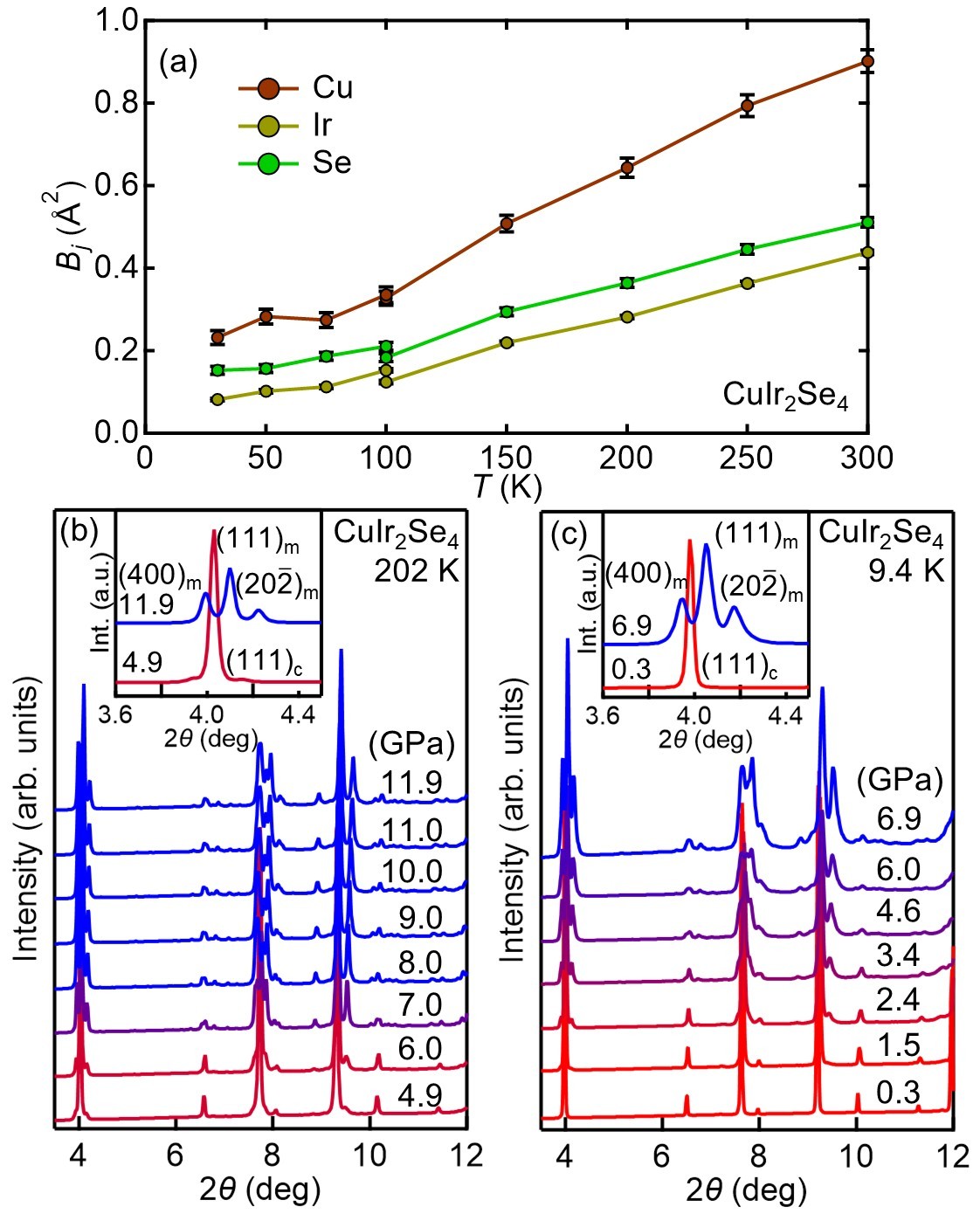}
\caption{\label{fig:Figure1} (a) Temperature dependence of the atomic displacement parameters obtained from ambient-pressure powder X-ray diffraction measurements for CuIr$_2$Se$_4$. (b, c) Pressure dependence of the synchrotron X-ray diffraction patterns of CuIr$_2$Se$_4$ at (b) 202 K and (c) 9.4 K. The insets highlight characteristic changes in selected reflections across the structural transition. Reflections indexed in the low-pressure cubic and high-pressure monoclinic settings are labeled with the subscripts $c$ and $m$, respectively. The transition occurs at a lower pressure at 9.4 K than at 202 K. }
\end{figure}

We first examined the temperature evolution of the average crystal structure of CuIr$_2$Se$_4$ at ambient pressure using powder X-ray diffraction. As shown in Fig.~\ref{fig:Figure1}(a), the atomic displacement parameters obtained from the refinements remain within physically reasonable ranges for all constituent atoms and decrease smoothly upon cooling. No anomalous temperature dependence is detected within the experimental resolution. Thus, the present powder-diffraction results provide no evidence of a crystallographic phase transition in the average structure of CuIr$_2$Se$_4$ over the measured temperature range.

Figures~\ref{fig:Figure1}(b) and \ref{fig:Figure1}(c) show the pressure dependences of the diffraction patterns of CuIr$_2$Se$_4$ at 202~K and 9.4~K, respectively. At both temperatures, Bragg-peak splitting and the emergence of superlattice reflections are observed with increasing pressure, demonstrating a pressure-induced structural transition from the low-pressure cubic phase to a high-pressure phase.

To characterize the high-pressure phase, we performed Le Bail analyses of its diffraction patterns. The reflections could be indexed using a monoclinic unit cell, and the observed systematic absences narrowed the possible space groups to $C2/c$ and $Cc$. Hereafter, we therefore refer to the low- and high-pressure phases as the cubic and monoclinic phases, respectively. The insets of Figs.~\ref{fig:Figure1}(b) and \ref{fig:Figure1}(c) highlight the characteristic changes in selected reflections across the transition. Reflections indexed in the cubic and monoclinic settings are labeled with the subscripts $c$ and $m$, respectively. The same subscripts are used below to distinguish lattice parameters and other crystallographic quantities associated with the two phases. The onset of the diffraction changes occurs at a lower pressure at 9.4~K than at 202~K. This observation is consistent with a positive slope of the low-pressure boundary of the monoclinic-phase stability region in the pressure--temperature phase diagram; that is, on the low-pressure side, the structural transition temperature increases with increasing pressure.

\begin{figure*}
\includegraphics[width=170mm]{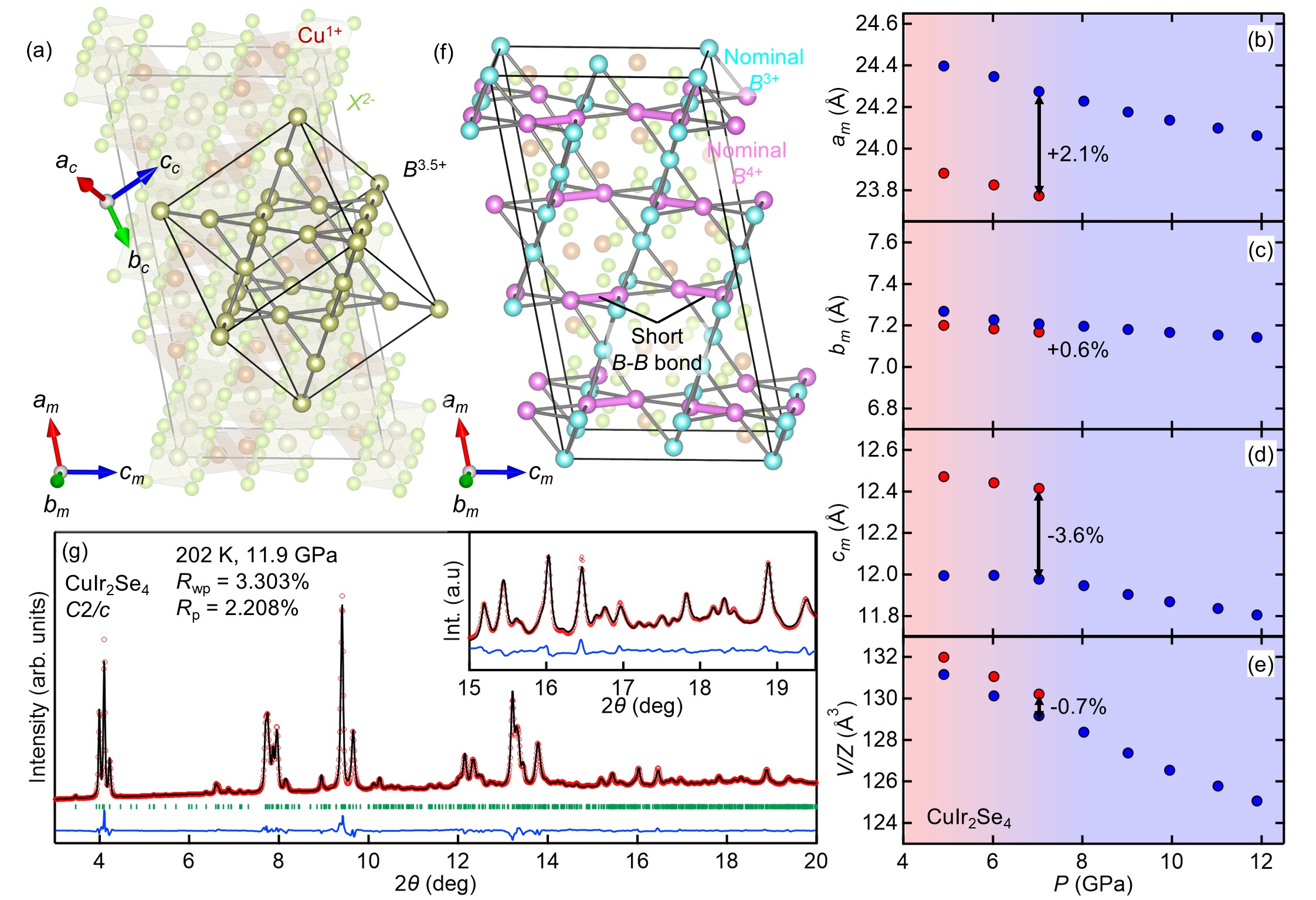}
\caption{\label{fig:Figure2} Structural analysis of the high-pressure phase of CuIr$_2$Se$_4$. For comparison with the Rh-based compounds discussed below, the transition-metal and chalcogen sites are denoted generically as $B$ and $X$, respectively; the quantitative structural model and diffraction fit shown here pertain to CuIr$_2$Se$_4$. (a) Relationship between the low-pressure cubic structure and the monoclinic unit-cell setting. The $B$-site pyrochlore sublattice is highlighted, and the cubic structure is shown semi-transparently in the monoclinic setting. Within this setting, kagome and triangular $B$-site layers alternate along the $a_m$ direction. (b--e) Pressure dependences of the monoclinic lattice parameters $a_m$, $b_m$, and $c_m$, and the unit-cell volume per formula unit $V/Z$. Red and blue circles represent the cubic and monoclinic phases, respectively. The lattice parameters of the cubic phase were converted to the corresponding monoclinic setting using $a_m=\sqrt{11/2}\,a_c$, $b_m=\sqrt{1/2}\,b_c$, and $c_m=\sqrt{3/2}\,c_c$. The unit-cell volume is normalized by the number of formula units $Z$. (f) DFT-relaxed Phase-IV-type structural model of the high-pressure phase of CuIr$_2$Se$_4$. The fractional atomic coordinates of this model were used without further refinement in the constrained profile fit shown in panel (g). The $B$ sites participating in the short $B$--$B$ bonds and the remaining $B$ sites are shown in pink and light blue, respectively. The corresponding nominal $B^{4+}$ and $B^{3+}$ labels are used solely as a bookkeeping scheme to represent the short-bond topology within an idealized ionic picture and do not imply complete integer charge separation. The short $B$--$B$ bonds are indicated by thick pink lines, while the Cu and $X$ sites are shown semi-transparently. (g) Constrained profile fit to the synchrotron powder X-ray diffraction data for CuIr$_2$Se$_4$ collected at 202~K and 11.9~GPa, using the structural model shown in panel (f). The fractional atomic coordinates were fixed to those of the DFT-relaxed model, whereas the lattice parameters, scale factor, background, and profile parameters were refined. Panels (f) and (g) therefore share the same fractional atomic coordinates.}
\end{figure*}

The relationship between the cubic and monoclinic unit cells is illustrated in Fig.~\ref{fig:Figure2}(a). The monoclinic cell has the same supercell relationship to the parent cubic spinel lattice as the Phase-IV cell experimentally established for CuIr$_2$S$_4$ \cite{CuIr2S4_emi}. Within this monoclinic setting, the Ir pyrochlore sublattice can be viewed as alternating kagome and triangular layers stacked along the $a_m$ direction, with the kagome layers lying in the $b_m$--$c_m$ plane. The unit-cell volume of the monoclinic phase is approximately twice that of the low-pressure cubic phase, resulting in a complex structure containing a large number of crystallographically independent atomic sites. Furthermore, because the reflections in the DAC powder-diffraction data overlap substantially and the X-ray scattering is dominated by Ir, it was difficult to refine all atomic coordinates simultaneously as independent free parameters.

We therefore first focused on the unit-cell geometry and its pressure evolution. The pressure dependences of the lattice parameters are shown in Figs.~\ref{fig:Figure2}(b)--\ref{fig:Figure2}(e). Upon the structural phase transition, $a_m$ increases and $c_m$ decreases, whereas the change in $b_m$ is comparatively small. This anisotropic lattice response is consistent with the formation of Ir--Ir short bonds primarily within the kagome layers in the $b_m$--$c_m$ plane. The same qualitative lattice response was observed in Phase IV of CuIr$_2$S$_4$ \cite{CuIr2S4_emi}, further supporting a close structural correspondence between the two high-pressure phases.

Guided by this correspondence, we applied the structural-analysis framework developed for Phase IV of CuIr$_2$S$_4$ \cite{CuIr2S4_emi}. Selected coordinates of the Ir atoms in the kagome layers were first refined to construct a diffraction-informed starting model and identify a candidate short-bond topology. This model was subsequently used as the initial structure for DFT-based structural optimization, in which the internal coordinates of all constituent atoms, including Cu and Se, were optimized. Structural optimization initiated from the $Cc$ model converged to $C2/c$ symmetry. The resulting DFT-relaxed model is shown in Fig.~\ref{fig:Figure2}(f), and Fig.~\ref{fig:Figure2}(g) shows a constrained profile fit to the diffraction data collected at 202 K and 11.9 GPa. The fractional atomic coordinates were fixed to those of the DFT-relaxed model shown in panel (f), whereas the lattice parameters, scale factor, background, and profile parameters were refined. Thus, panels (f) and (g) share the same fractional atomic coordinates, and panel (g) evaluates the compatibility of the DFT-relaxed model with the observed diffraction pattern rather than providing an independent refinement of the atomic positions. The model reproduces both the fundamental and weak superlattice reflections, showing that it is compatible with the principal diffraction features of the high-pressure phase.
 
In the DFT-relaxed model, selected Ir--Ir distances are shortened to values comparable to those associated with Ir--Ir dimers in Phases II and IV of CuIr$_2$S$_4$, supporting Phase-IV-type bond disproportionation. The Ir sites participating in these short bonds account for one-half of the Ir sites in the unit cell. To describe this short-bond topology within an idealized ionic picture, we assign nominal Ir$^{3+}$ and Ir$^{4+}$ labels solely as a bookkeeping scheme. The sites participating in the short Ir--Ir bonds are labeled nominal Ir$^{4+}$, whereas the remaining sites are labeled nominal Ir$^{3+}$. This 1:1 assignment is consistent with the average Ir valence of $+3.5$, but does not imply complete integer charge separation. Within this nominal-valence representation, the DFT-relaxed model exhibits the same short-bond topology as that found in Phase IV of CuIr$_2$S$_4$ \cite{CuIr2S4_emi}. Thus, the diffraction data are consistent with CuIr$_2$Se$_4$ adopting a Phase-IV-type bond-disproportionated structure under pressure.

\subsection{High-pressure structural evolution of CuRh$_2$S$_4$ and CuRh$_2$Se$_4$}

\begin{figure*}
\includegraphics[width=170mm]{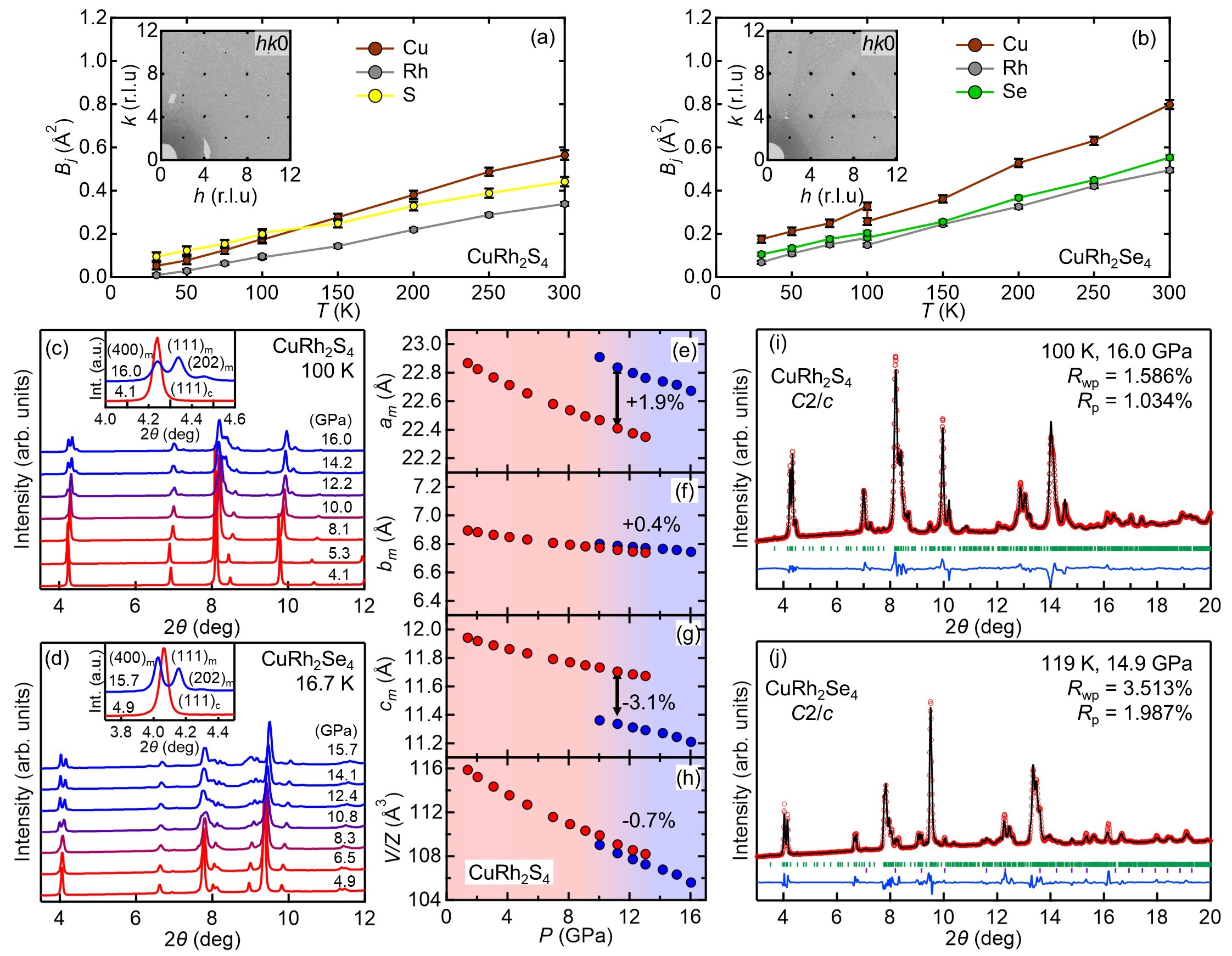}
\caption{\label{fig:Figure3} (a, b) Temperature dependence of the atomic displacement parameters obtained from ambient-pressure powder X-ray diffraction measurements for (a) CuRh$_2$S$_4$ and (b) CuRh$_2$Se$_4$. The insets show the single-crystal X-ray diffraction patterns measured at 100 K in the \textit{hk}0 plane. Here, the subscripts $c$ and $m$ attached to the Miller indices denote the low-pressure cubic and high-pressure monoclinic settings, respectively. (c, d) Pressure dependences of the synchrotron X-ray diffraction patterns of (c) CuRh$_2$S$_4$ at 100 K and (d) CuRh$_2$Se$_4$ at 16.7 K. The insets highlight the characteristic changes in selected reflections, indicating a lowering of symmetry. Although significant peak broadening develops in CuRh$_2$Se$_4$ under pressure, the splitting pattern is essentially the same as that observed in CuIr$_2$Se$_4$ and CuRh$_2$S$_4$, suggesting the formation of a similar symmetry-lowered phase. (e--h) Pressure dependences of the lattice parameters $a_m$, $b_m$, and $c_m$, and the unit-cell volume per formula unit \textit{V/Z} of CuRh$_2$S$_4$. The lattice parameters of the cubic phase were converted to the corresponding monoclinic setting using $a_m=\sqrt{11/2}\,a_c$, $b_m=\sqrt{1/2}\,b_c$, and
$c_m=\sqrt{3/2}\,c_c$. The unit-cell volume is normalized by the number of formula units \textit{Z}, so that the low-pressure cubic and high-pressure monoclinic phases can be compared on the same basis. (i) Constrained profile fit to the synchrotron powder X-ray diffraction data for CuRh$_2$S$_4$ collected at 100~K and 16.0~GPa, using the DFT-relaxed $C2/c$ model. The fractional atomic coordinates were fixed to those of the DFT-relaxed model, and the atomic displacement parameters were fixed to the corresponding ambient-pressure values, whereas the lattice parameters, scale factor, background, and profile parameters were refined. The fit supports the compatibility of CuRh$_2$S$_4$ with a Phase-IV-type structural model rather than providing an independent refinement of the atomic positions. (j) Le Bail fit to the synchrotron powder X-ray diffraction data for CuRh$_2$Se$_4$ collected at 119~K and 14.9~GPa using a $C$-centered monoclinic cell. Rietveld refinements using the candidate structural model and preferred-orientation corrections did not yield reliable structural parameters; therefore, only the Le Bail fit is presented. The cell has the same supercell relationship to the parent cubic structure as the high-pressure cells of CuIr$_2$Se$_4$ and CuRh$_2$S$_4$. This analysis establishes a compatible monoclinic supercell, although the atomic-scale ordering pattern remains unresolved.}
\end{figure*}

Figures~\ref{fig:Figure3}(a) and \ref{fig:Figure3}(b) show the temperature dependences of the atomic displacement parameters of CuRh$_2$S$_4$ and CuRh$_2$Se$_4$, respectively, obtained from powder-diffraction refinements at ambient pressure. For both compounds, the atomic displacement parameters remain within physically reasonable ranges for all constituent atoms, including Rh, and decrease smoothly upon cooling, with no anomalous temperature dependence. The insets of Figs.~\ref{fig:Figure3}(a) and \ref{fig:Figure3}(b) show synchrotron single-crystal X-ray diffraction patterns measured at 100~K in the $hk0$ plane. Both compounds exhibit sharp Bragg reflections, with no evident diffuse streaks in the measured plane. Thus, the powder-diffraction results provide no evidence of a crystallographic phase transition in the average structures over the investigated temperature range, while the single-crystal data reveal no obvious signatures of crystallographic disorder at 100~K.

Against this ambient-pressure baseline, pressure-induced structural transitions are clearly observed in both CuRh$_2$S$_4$ and CuRh$_2$Se$_4$, as shown in Figs.~\ref{fig:Figure3}(c) and \ref{fig:Figure3}(d). In both compounds, increasing pressure produces a clear splitting of Bragg reflections, marking a transition from the low-pressure cubic phase to a lower-symmetry high-pressure phase. The Rh-based compounds exhibit more pronounced peak broadening under pressure than CuIr$_2$Se$_4$, particularly in the case of CuRh$_2$Se$_4$, which limits the precision of the subsequent structural analysis. Nevertheless, the characteristic evolution of the diffraction patterns resembles that observed in the high-pressure phases of CuIr$_2$S$_4$ and CuIr$_2$Se$_4$, suggesting that the resulting symmetry-lowered phases have closely related crystallographic frameworks.

For CuRh$_2$S$_4$, the high-pressure reflections could be indexed using the same monoclinic supercell setting as that used for CuIr$_2$Se$_4$. Figures~\ref{fig:Figure3}(e)--\ref{fig:Figure3}(h) show the pressure dependences of the monoclinic lattice parameters and the unit-cell volume per formula unit. The high-pressure phase exhibits an anisotropic lattice distortion similar to those observed in CuIr$_2$S$_4$ and CuIr$_2$Se$_4$. We therefore applied the same structural-analysis procedure as that used for CuIr$_2$Se$_4$ and obtained a DFT-relaxed candidate model with $C2/c$ symmetry. Figure~\ref{fig:Figure3}(i) shows a constrained profile fit to the diffraction data collected at 100~K and 16.0~GPa. The fractional atomic coordinates were fixed to those of the DFT-relaxed model, and the atomic displacement parameters were fixed to the corresponding ambient-pressure values, whereas the lattice parameters, scale factor, background, and profile parameters were refined. Thus, this fit evaluates the compatibility of the DFT-relaxed model with the observed diffraction pattern rather than providing an independent refinement of the atomic positions. In this model, approximately half of the Rh sites in the kagome layers participate in short Rh--Rh bonds, producing the same short-bond topology as that found in the Phase-IV-type models of CuIr$_2$S$_4$ and CuIr$_2$Se$_4$. The diffraction data are therefore compatible with a Phase-IV-type bond-disproportionated structure in CuRh$_2$S$_4$.

For CuRh$_2$Se$_4$, the diffraction peaks could be indexed and the monoclinic lattice parameters were obtained by Le Bail fitting. Rietveld refinements based on the candidate monoclinic structural model were also attempted with preferred-orientation corrections. However, even after applying the preferred-orientation corrections, the calculated profiles did not adequately reproduce the observed relative intensities, and the refinements did not yield stable or physically meaningful structural parameters. The difficulty is attributed primarily to the severe preferred orientation of the compressed powder, together with the pronounced diffraction-peak broadening and overlap. Because the Le Bail method treats the reflection intensities independently, these effects do not prevent evaluation of the lattice metrics, whereas they preclude a reliable intensity-based determination of the atomic coordinates and short-bond topology. Nevertheless, the integrated intensity of the cubic (111) reflection can be used to track the residual cubic-phase content semiquantitatively. We therefore use this single-reflection measure only to follow the progress of the structural transformation and not to determine the atomic structure of the high-pressure phase. Accordingly, we restrict our conclusion for CuRh$_2$Se$_4$ to the identification of a compatible monoclinic supercell; the present data neither confirm nor exclude a Phase-IV-type atomic arrangement.

Overall, the diffraction results establish closely related monoclinic high-pressure supercells in all three compounds. For CuIr$_2$Se$_4$ and CuRh$_2$S$_4$, constrained profile fits using DFT-relaxed models support compatibility with Phase-IV-type bond-disproportionated structures. For CuRh$_2$Se$_4$, the available data establish the corresponding monoclinic supercell, but not its atomic-scale ordering pattern. The material-dependent pressure and temperature ranges over which these structural transformations proceed are compared in the following section.

\subsection{Pressure–temperature phase diagrams and electronic properties}

\begin{figure*}
\includegraphics[width=170mm]{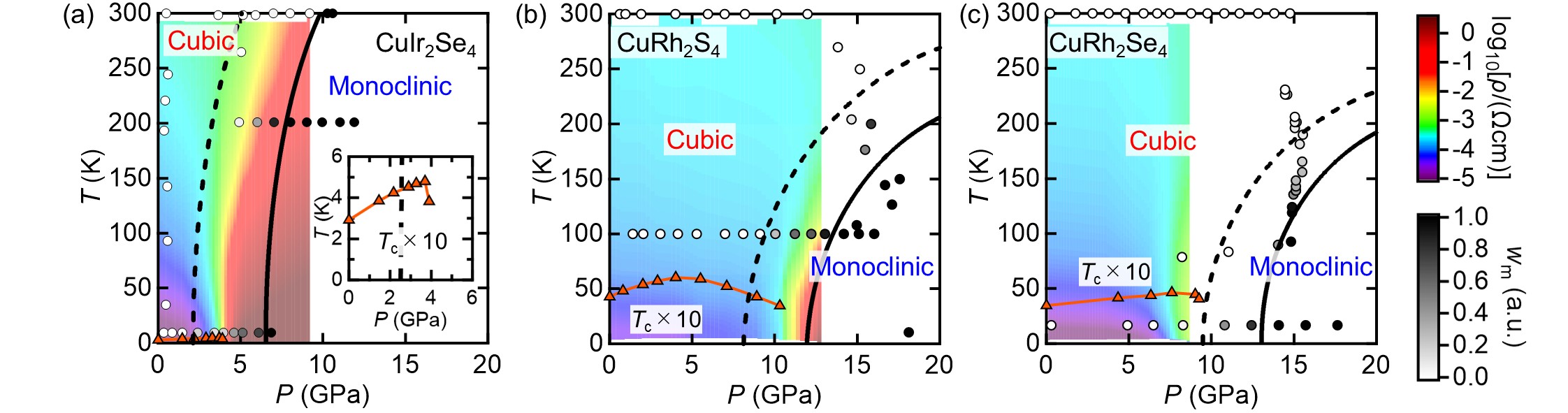}
\caption{\label{fig:Figure4} Pressure--temperature phase diagrams of (a) CuIr$_2$Se$_4$, (b) CuRh$_2$S$_4$, and (c) CuRh$_2$Se$_4$. The color maps show log$_{10}[\rho/(\Omega\mathrm{cm})]$, constructed from the electrical-resistivity data shown in Figs.~\ref{fig:Figure7}(a)--\ref{fig:Figure7}(c). The color shading was generated by linear interpolation between the measured data points, and the corresponding resistivity scale is shown by the upper color bar. The grayscale of the circles represents the relative monoclinic content $w_m$: white and black correspond to $w_m=0$ (cubic phase) and $w_m=1$ (monoclinic phase), respectively, while intermediate gray levels indicate coexistence of the two phases. The corresponding $w_m$ scale is shown by the lower color bar. For CuIr$_2$Se$_4$ and CuRh$_2$S$_4$, $w_m$ was obtained from two-phase profile refinements using RIETAN-FP. For CuRh$_2$Se$_4$, the relative cubic-phase content $w_c$ was estimated from the integrated-intensity ratio of the cubic (111) reflection, and $w_m$ was defined as $1-w_c$. The $w_m$ values for CuRh$_2$Se$_4$ thus provide a semiquantitative measure of the progress of the structural transformation rather than absolute phase fractions. The dashed and solid curves are guides to the eye indicating the onset and completion, respectively, of the structural transformation with increasing pressure. The low- and high-pressure single-phase regions are labeled cubic and monoclinic, respectively, and the region between the dashed and solid curves represents coexistence of the two phases. Triangles represent the zero-resistance superconducting transition temperature $T_{\mathrm{c}}$ and are plotted at $T_{\mathrm{c}}$$\times$10 on the temperature axis for visibility. The inset in panel (a) shows an enlarged view of the pressure dependence of $T_{\mathrm{c}}$$\times$10 for CuIr$_2$Se$_4$.
}
\end{figure*}

Diffraction experiments were performed under multiple pressure and temperature conditions for CuIr$_2$Se$_4$, CuRh$_2$S$_4$, and CuRh$_2$Se$_4$, and the resulting pressure--temperature phase diagrams are summarized in Figs.~\ref{fig:Figure4}(a)--\ref{fig:Figure4}(c). The resistivity maps and the pressure evolution of superconductivity are also shown in this figure and are discussed in the subsections below. At each pressure--temperature point, the relative monoclinic content $w_m$ is represented by the grayscale of the corresponding symbol. For CuIr$_2$Se$_4$ and CuRh$_2$S$_4$, $w_m$ was obtained from two-phase profile refinements, whereas for CuRh$_2$Se$_4$ it was estimated from the integrated-intensity ratio of the cubic (111) reflection, as described in Sec.~II. In these panels, the dashed curve indicates the onset of the monoclinic phase, whereas the solid curve marks the completion of the structural transformation. This representation allows us to visualize not only the phase boundary itself but also the width of the transformation region between the cubic and monoclinic phases.

A two-phase coexistence region is observed in all three compounds during the transformation from the low-pressure cubic phase to the high-pressure monoclinic phase. The coexistence of distinct cubic and monoclinic diffraction profiles is consistent with a first-order structural transition, although the finite experimental width of the coexistence region need not represent an equilibrium two-phase field, as discussed in Sec.~III D. At the same time, the breadth and overall shape of the transformation region show clear material dependence.

\subsubsection{Ambient-pressure superconductivity and electronic structure}

\begin{figure}
\includegraphics[width=85mm]{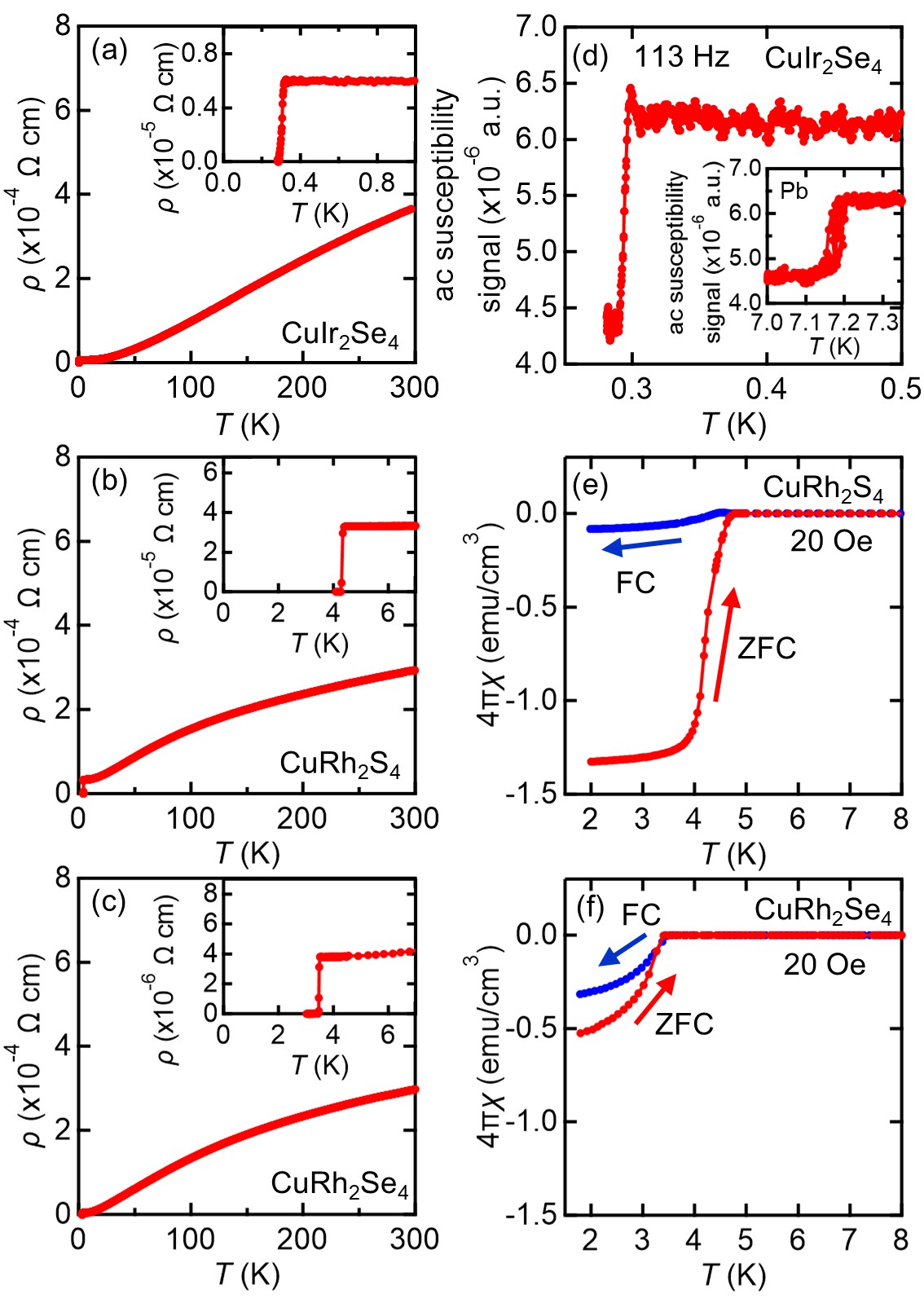}
\caption{\label{fig:Figure5} Ambient-pressure superconducting properties of (a,d) CuIr$_2$Se$_4$, (b,e) CuRh$_2$S$_4$, and (c,f) CuRh$_2$Se$_4$. (a--c) Temperature dependences of the electrical resistivity at ambient pressure. The insets show enlarged views of the low-temperature region around the superconducting transitions. (d) Temperature dependence of the ac magnetic-susceptibility signal of CuIr$_2$Se$_4$ measured at 113~Hz. The inset shows the corresponding signal from a Pb reference with a volume comparable to that of the CuIr$_2$Se$_4$ specimen, measured under identical conditions. The signal changes in the main panel and inset are plotted using the same vertical scale. (e,f) Temperature dependences of the dc magnetic susceptibility of CuRh$_2$S$_4$ and CuRh$_2$Se$_4$, respectively, measured using zero-field-cooled (ZFC) and field-cooled (FC) protocols. The applied dc magnetic fields are indicated in the respective panels. All three compounds exhibit zero resistance accompanied by corresponding diamagnetic responses. For CuIr$_2$Se$_4$, comparison with the Pb reference indicates a superconducting shielding fraction close to 100\%. The zero-resistance superconducting transition temperatures $T_c$ are 0.29~K for CuIr$_2$Se$_4$, 4.27~K for CuRh$_2$S$_4$, and 3.45~K for CuRh$_2$Se$_4$. The bulk superconductivity observed in CuIr$_2$Se$_4$ has not been reported previously.
}
\end{figure}

Figure~\ref{fig:Figure5} summarizes the ambient-pressure superconducting properties of the three compounds. All three exhibit metallic conduction above their superconducting transitions and enter a superconducting state at low temperature. For CuRh$_2$S$_4$ and CuRh$_2$Se$_4$, the observation of zero resistance together with corresponding diamagnetic responses confirms superconductivity in the present samples, consistent with previous reports \cite{CuRh2S4_SC1,CuRh2Se4_SC1,CuRh2S4_CuRh2Se4_SC1,CuRh2S4_CuRh2Se4_SC2,CuRh2S4_CuRh2Se4_SC3,CuRh2S4_CuRh2Se4_Heat}. The zero-resistance transition temperatures are 4.27~K for CuRh$_2$S$_4$ and 3.45~K for CuRh$_2$Se$_4$.

Notably, we establish bulk superconductivity in CuIr$_2$Se$_4$ at ambient pressure. As shown in Figs.~\ref{fig:Figure5}(a) and \ref{fig:Figure5}(d), the electrical resistivity reaches zero at 0.29~K, and the zero-resistance transition was reproduced in a second specimen. The ac magnetic susceptibility measured at 113~Hz exhibits a clear diamagnetic response at approximately the same temperature. The magnitude of the susceptibility change was compared with that of a Pb reference having a comparable volume and measured under identical conditions. This comparison indicates a superconducting shielding fraction close to 100\%. The reproducible zero resistance and nearly complete diamagnetic shielding together demonstrate that superconductivity in CuIr$_2$Se$_4$ is a bulk property. Thus, although all three compounds are bulk superconductors at ambient pressure, CuIr$_2$Se$_4$ is distinguished by its substantially lower superconducting transition temperature.

\begin{figure*}
\includegraphics[width=170mm]{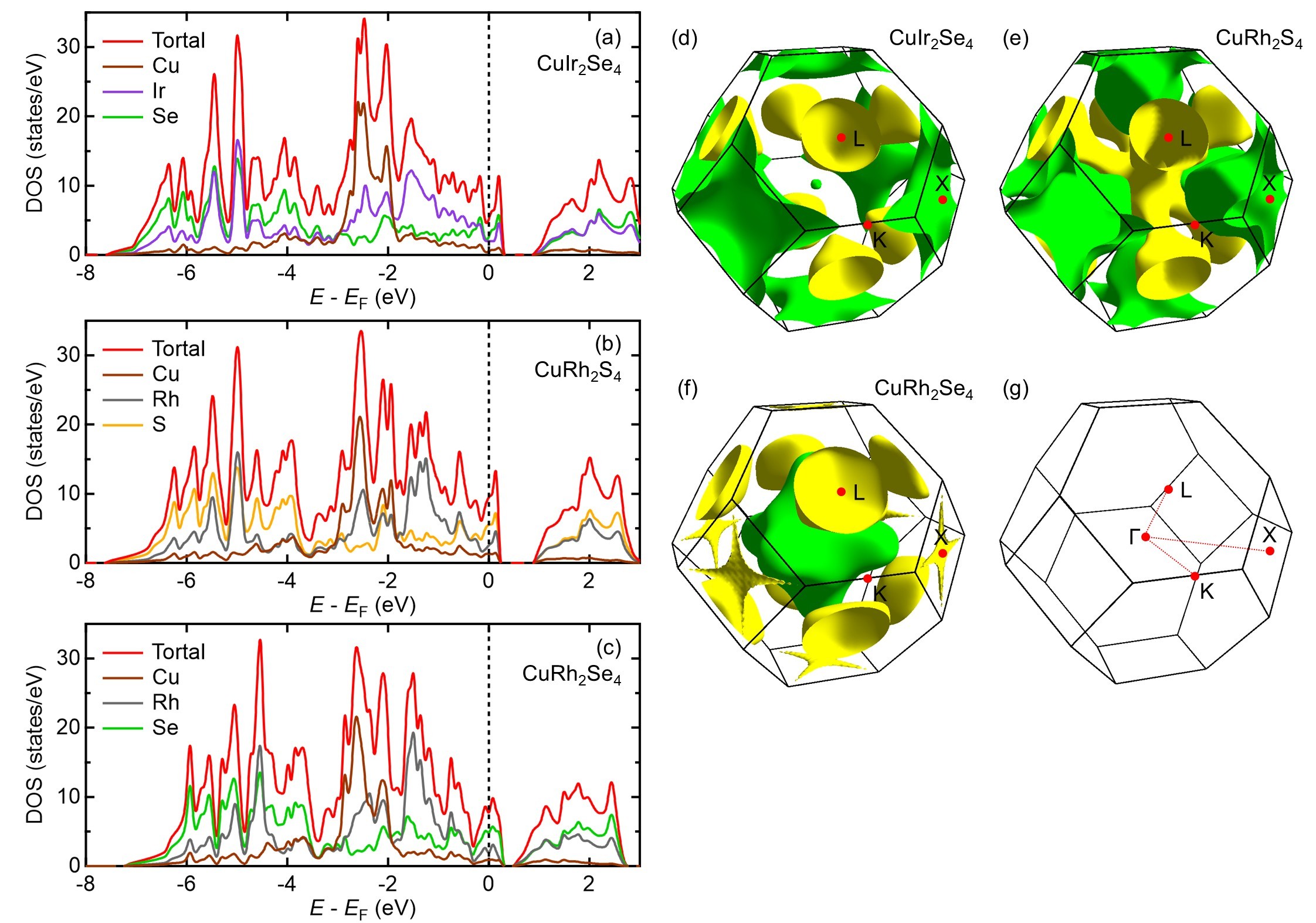}
\caption{\label{fig:Figure6} Density of states (DOS) and Fermi surfaces of the ambient-pressure cubic phases obtained from DFT calculations including spin--orbit coupling. The calculations were performed using fully optimized structures with $Fd\bar{3}m$ symmetry. (a--c) Total and atom-projected DOS of (a) CuIr$_2$Se$_4$, (b) CuRh$_2$S$_4$, and (c) CuRh$_2$Se$_4$, plotted as a function of energy relative to the Fermi level, $E-E_{\mathrm{F}}$. The dashed vertical lines at $E-E_{\mathrm{F}}=0$ indicate the Fermi level. (d--f) Corresponding Fermi surfaces within the first Brillouin zone for (d) CuIr$_2$Se$_4$, (e) CuRh$_2$S$_4$, and (f) CuRh$_2$Se$_4$. The white lines outline the first Brillouin zone, and different colors are used to distinguish the Fermi-surface sheets. All three Fermi surfaces are shown in the same crystallographic orientation. A $\Gamma$-centered Fermi-surface sheet is present in CuRh$_2$S$_4$ and CuRh$_2$Se$_4$ but absent in CuIr$_2$Se$_4$. (g) First Brillouin zone of the cubic phase showing the high-symmetry points indicated in panels (d)--(f).}
\end{figure*}

Figure~\ref{fig:Figure6} compares the calculated density of states (DOS) and Fermi surfaces of the ambient-pressure cubic phases. The calculated DOS at the Fermi level, $N(E_{\mathrm{F}})$, is approximately 5.2 states~eV$^{-1}$ for CuIr$_2$Se$_4$, 7.7 states~eV$^{-1}$ for CuRh$_2$S$_4$, and 9.5 states~eV$^{-1}$ for CuRh$_2$Se$_4$. Thus, $N(E_{\mathrm{F}})$ is substantially smaller in CuIr$_2$Se$_4$ than in the Rh-based compounds. The atom-projected DOS shows that the states near $E_{\mathrm{F}}$ are dominated by the transition-metal $d$ and chalcogen $p$ orbitals, whereas the Cu-derived states are concentrated mainly between approximately $-3$ and $-2$~eV.

The calculated Fermi surfaces reveal a notable difference in topology: a Fermi-surface sheet centered at the $\Gamma$ point is present in CuRh$_2$S$_4$ and CuRh$_2$Se$_4$ but absent in CuIr$_2$Se$_4$. A comparison of the two selenides is particularly informative. Their Fermi-surface sheets around the $L$ points are similar in size, suggesting that these sheets are unlikely to be the primary origin of the difference in $N(E_{\mathrm{F}})$. Instead, the principal contrast lies between the sizable $\Gamma$-centered sheet in CuRh$_2$Se$_4$ and the $X$-centered sheet in CuIr$_2$Se$_4$. CuRh$_2$Se$_4$ also possesses an extremely thin $X$-centered sheet, which is expected to make only a minor contribution to $N(E_{\mathrm{F}})$. These differences qualitatively associate the local depression of the DOS at $E_{\mathrm{F}}$ in CuIr$_2$Se$_4$ with changes in the $\Gamma$- and $X$-centered Fermi-surface sheets, rather than with a uniform reduction of the DOS contributions from all bands. Because the contributions of the individual Fermi-surface sheets to $N(E_{\mathrm{F}})$ have not been evaluated quantitatively, however, the contribution associated with the absence of the $\Gamma$-centered sheet cannot be specified as a numerical value or fraction.

Although the superconducting transition temperature is not determined by $N(E_{\mathrm{F}})$ alone, the reduced $N(E_{\mathrm{F}})$ of CuIr$_2$Se$_4$ may be one factor contributing to its substantially lower $T_{\mathrm{c}}$. At the same time, the difference in $T_{\mathrm{c}}$ between CuRh$_2$S$_4$ and CuRh$_2$Se$_4$ does not follow their calculated $N(E_{\mathrm{F}})$ values directly, indicating that the DOS and Fermi-surface topology alone cannot account for the material dependence of superconductivity. These ambient-pressure electronic-structure differences provide a reference point for examining the distinct pressure responses discussed below.

\subsubsection{Pressure evolution of resistivity and superconductivity}

\begin{figure}
\includegraphics[width=85mm]{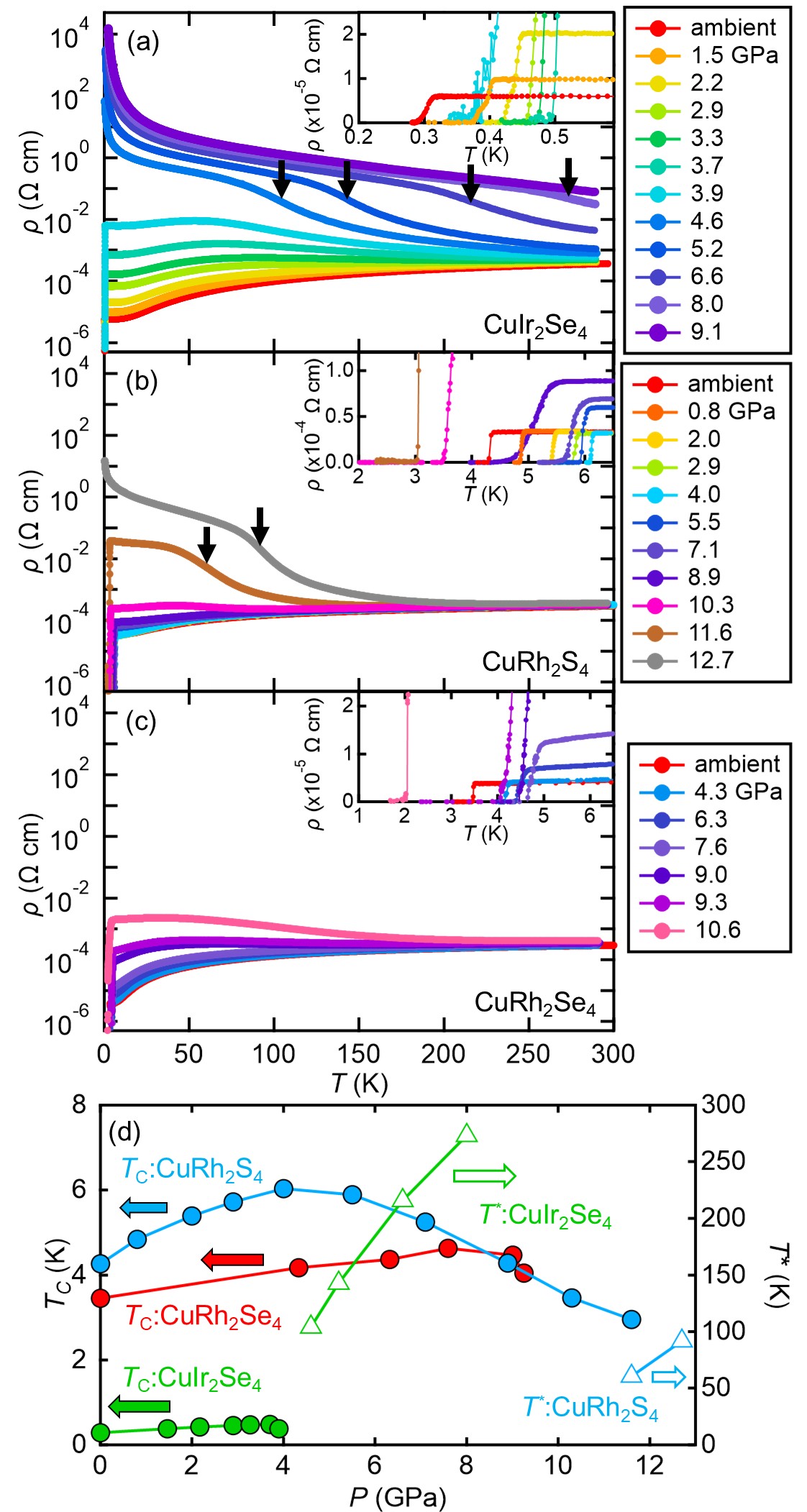}
\caption{\label{fig:Figure7} (a--c) Temperature dependences of the electrical resistivity of (a) CuIr$_2$Se$_4$, (b) CuRh$_2$S$_4$, and (c) CuRh$_2$Se$_4$ under pressure. The insets show enlarged views of the low-temperature region, highlighting the pressure evolution of the superconducting transition. In all three compounds, the resistivity increases with pressure and an insulating temperature dependence develops at high pressures, although the magnitude and sharpness of this evolution are strongly material dependent. The downward arrows in panels (a) and (b) indicate $T^*$, determined from the minimum in $d\ln\rho/dT$. No corresponding $T^*$ was assigned to CuRh$_2$Se$_4$ because the minimum-like feature at 10.6 GPa is shallow and broad and is accompanied by only a weak insulating-like upturn in $\rho$($T$). (d) Pressure dependences of the superconducting transition temperature $T_{\mathrm{c}}$ and the characteristic crossover temperature $T^{*}$ for CuIr$_2$Se$_4$, CuRh$_2$S$_4$, and CuRh$_2$Se$_4$. Circles and triangles represent $T_{\mathrm{c}}$ and $T^{*}$, respectively. The value of $T_{\mathrm{c}}$ is defined as the temperature at which the electrical resistivity reaches zero, whereas $T^{*}$ is defined as the temperature corresponding to the minimum in $d\ln\rho/dT$ near the insulating crossover.
}
\end{figure}

The results of high-pressure electrical resistivity measurements for CuIr$_2$Se$_4$, CuRh$_2$S$_4$, and CuRh$_2$Se$_4$ are shown in Figs.~\ref{fig:Figure7}(a)--\ref{fig:Figure7}(c). The measurements were performed up to 9.1 GPa for CuIr$_2$Se$_4$, 12.7 GPa for CuRh$_2$S$_4$, and 10.6 GPa for CuRh$_2$Se$_4$, thereby extending the pressure range beyond the previously reported values of 5.0, 8.0, and 6.5 GPa, respectively \cite{CuIr2Se4_hp_cond_1,CuRh2S4_hp_con_XRD,CuRh2Se4_hp_con1}. This extended pressure coverage enables a systematic comparison of how pressure-induced insulating transport develops in the three compounds.

At low pressures, all three compounds exhibit metallic transport. With increasing pressure, the resistivity increases and an insulating-like temperature dependence, $d\rho/dT<0$, develops upon cooling. The temperature dependences of $d\ln\rho/dT$ are shown in Appendix C. At intermediate pressures, the resistivity curves exhibit a broad crossover with an inflection point as a function of temperature rather than a sharp transition. This behavior is consistent with the two-phase coexistence regions between the low-pressure cubic and high-pressure monoclinic phases identified by diffraction, as discussed below. The broad resistive response can therefore be understood, at least in part, as arising from the progressive change in the relative phase fractions across the coexistence region. Because electrical transport through a mixed-phase polycrystalline specimen can be percolative, however, the resistivity is not expected to vary linearly with the structural phase fraction. Thus, a pressure-driven evolution from metallic toward insulating transport is a common feature of this series.

The onset pressure and subsequent evolution of the insulating-like behavior, however, differ markedly among the compounds. In CuIr$_2$Se$_4$, insulating-like transport begins to develop at a relatively low pressure and rapidly extends over a broad temperature range with further compression. At 9.1 GPa, $d\rho/dT$ is negative over the entire measured temperature range up to 300~K. In CuRh$_2$S$_4$, the onset occurs at a considerably higher pressure, and metallic transport persists near room temperature even at the highest pressure investigated. In CuRh$_2$Se$_4$, pressure enhances the resistivity, but the insulating-like upturn upon cooling remains only partially developed within the present pressure range. Still higher pressures would therefore likely be required to establish insulating behavior over a temperature range comparable to those observed in CuIr$_2$Se$_4$ and CuRh$_2$S$_4$.

To characterize the pressure-induced insulating crossover quantitatively, we analyzed the temperature derivative of the electrical resistivity. Because the resistive anomaly extends over a broad temperature range, consistent with the finite two-phase coexistence region identified by diffraction, a unique transition temperature cannot be determined directly from the raw resistivity curves. When a well-defined minimum associated with the insulating crossover is present, we define the characteristic crossover temperature $T^{*}$ as the temperature of this minimum in $d\ln\rho/dT$. Since $d\ln\rho/dT$ represents the relative change in resistivity per unit temperature, this minimum corresponds to the temperature at which the resistivity increases most rapidly upon cooling. The pressure dependence of $T^{*}$ is shown in Fig.~\ref{fig:Figure7}(d).

A well-defined minimum in $d\ln\rho/dT$ is observed from 4.6~GPa in CuIr$_2$Se$_4$ and from 11.6~GPa in CuRh$_2$S$_4$. In CuIr$_2$Se$_4$, $T^{*}$ increases steeply with pressure and eventually exceeds the upper limit of the measured temperature range. In CuRh$_2$S$_4$, by contrast, $T^{*}$ increases more gradually. In CuRh$_2$Se$_4$, the $d\ln\rho/dT$ curve at 10.6 GPa exhibits a shallow and broad minimum-like feature. Because this feature is much less pronounced than those observed in CuIr$_2$Se$_4$ and CuRh$_2$S$_4$ and is accompanied by only a weak insulating-like upturn in $\rho$($T$), no value of $T^{*}$ is assigned. These results show that insulating-like transport emerges at a lower pressure and develops toward higher temperatures more rapidly in CuIr$_2$Se$_4$ than in the Rh-based compounds.

The relationship between $T^{*}$ and superconductivity also differs between CuIr$_2$Se$_4$ and CuRh$_2$S$_4$. In CuIr$_2$Se$_4$, the superconducting transition temperature $T_{\mathrm{c}}$ initially increases with pressure but decreases abruptly near the first appearance of $T^{*}$, and zero-resistance superconductivity is no longer observed at higher pressures. In CuRh$_2$S$_4$, $T_{\mathrm{c}}$ reaches a maximum at a pressure well below that at which $T^{*}$ first becomes identifiable and subsequently decreases more gradually. Zero-resistance superconductivity remains observable even within the pressure range where a finite $T^{*}$ is assigned. Thus, the development of insulating-like transport competes with superconductivity in both compounds, but the competition is abrupt in CuIr$_2$Se$_4$ and extends over a broader pressure range in CuRh$_2$S$_4$. Because no well-defined $T^{*}$ can be assigned to CuRh$_2$Se$_4$, the corresponding relationship cannot be evaluated on the same basis within the present pressure range.

Taken together, the high-pressure transport data reveal a pressure-driven evolution toward insulating transport in all three compounds, while the onset pressure, characteristic temperature scale, and relationship with superconductivity differ markedly among them. In the following subsection, we compare these transport results directly with the diffraction-derived phase diagrams to clarify their relationship with the pressure-induced structural transformations.

\subsubsection{Correspondence between transport and the diffraction-derived phase diagrams}
We now compare the transport properties with the diffraction-derived phase diagrams shown in Figs.~\ref{fig:Figure4}(a)--\ref{fig:Figure4}(c). In all three compounds, the pressure--temperature region in which the electrical resistivity begins to increase markedly overlaps with that in which the high-pressure monoclinic phase first appears. This correspondence supports a close coupling between the pressure-induced evolution toward insulating transport and the structural transformation, rather than their being independent instabilities.

The diffraction-derived evolution of the monoclinic content further clarifies this correspondence. Within the low-pressure cubic-phase region, the normal-state transport is metallic. As the system enters the coexistence region between the cubic and monoclinic phases, an insulating-like upturn begins to emerge in the resistivity. With further compression, the relative monoclinic content increases in parallel with a progressive enhancement of the insulating-like behavior.

The material dependence of the transport response also tracks the structural evolution. In CuIr$_2$Se$_4$, the transformation from the cubic to the monoclinic phase occurs over a relatively narrow pressure interval, and insulating-like transport develops rapidly with pressure. The pronounced increase in resistivity extending to room temperature is consistent with the appearance of the monoclinic phase at room temperature. In CuRh$_2$S$_4$ and CuRh$_2$Se$_4$, by contrast, the structural transformation extends over broader pressure ranges, accompanied by a more gradual development of insulating-like transport. Thus, the pressure evolution of the transport properties closely reflects the material-dependent breadth of the structural transformation.

Superconductivity exhibits a similarly material-dependent relationship with the structural phase diagrams. In CuIr$_2$Se$_4$, $T_{\mathrm{c}}$ initially increases with pressure but decreases abruptly as the monoclinic phase and insulating-like transport develop, and zero-resistance superconductivity is no longer observed at higher pressures. In CuRh$_2$S$_4$, $T_{\mathrm{c}}$ reaches a maximum before the insulating crossover becomes prominent and subsequently decreases more gradually, remaining finite within part of the structural-transition region. In CuRh$_2$Se$_4$, the pressure evolution of $T_{\mathrm{c}}$ is also gradual; however, because no well-defined $T^{*}$ can be assigned within the investigated pressure range, its relationship with the insulating crossover cannot be evaluated on the same basis. These contrasting behaviors show that superconductivity is progressively destabilized as the pressure-induced structural and transport anomalies develop, but that the manner of this competition depends strongly on the compound.

Taken together, the transport and diffraction results demonstrate a close correspondence between the formation of the monoclinic high-pressure phases and the development of insulating-like transport. The three compounds form closely related monoclinic supercells, although the atomic-scale ordering pattern of CuRh$_2$Se$_4$ remains unresolved. The material-dependent location and breadth of the structural-transition region correlate with both the rate at which insulating-like transport develops and the pressure evolution of superconductivity. These results support a strongly coupled structural and electronic transformation, while not by themselves establishing the microscopic direction of causality.

\subsection{Discussion}

To place the present results in context, it is useful first to clarify the terminology established for CuIr$_2$S$_4$. The ordered phase appearing at low temperature and ambient pressure is conventionally referred to as Phase II, whereas the distinct ordered phase stabilized under pressure is referred to as Phase IV \cite{CuIr2S4_emi}. The structure of Phase IV has been experimentally established for CuIr$_2$S$_4$, and the Phase-IV-type short-bond topology considered here is based on that structure. As shown in Fig.~\ref{fig:Figure2}(f), the DFT-relaxed model for CuIr$_2$Se$_4$ retains this topology. In the present compounds, the diffraction data for CuIr$_2$Se$_4$ and CuRh$_2$S$_4$ are compatible with DFT-relaxed Phase-IV-type structural models, whereas those for CuRh$_2$Se$_4$ establish a closely related monoclinic supercell but do not determine its atomic-scale ordering pattern. Thus, all three compounds exhibit a common tendency toward closely related monoclinic high-pressure supercells, although the level of structural information differs among them. By contrast, among these related compounds, only CuIr$_2$S$_4$ is known to realize a Phase-II-type state at low temperature and ambient pressure.

The experimentally established Phase-IV structure of CuIr$_2$S$_4$ and the supported Phase-IV-type models of CuIr$_2$Se$_4$ and CuRh$_2$S$_4$ are characterized by bond disproportionation: some transition-metal sites participate in short metal--metal bonds, whereas the remaining sites do not. If nominal Ir$^{3+}$/Ir$^{4+}$ or Rh$^{3+}$/Rh$^{4+}$ labels are assigned to this short-bond topology as a bookkeeping scheme, each tetrahedron contains two nominal 3+ and two nominal 4+ sites, consistent with the Anderson condition \cite{anderson,CsW2O6,LiRh2O4}. These labels do not imply complete crystallization of integer valences. Rather, they provide an idealized ionic representation of the local short-bond topology. The corresponding high-pressure structures are therefore more appropriately described as bond-disproportionated states consistent with an Anderson-type local tetrahedron rule than as simple integer-valence charge-ordered states. This atomic-scale assignment cannot yet be made for CuRh$_2$Se$_4$, for which only the compatible monoclinic supercell has been established.

Phase II of CuIr$_2$S$_4$, by contrast, appears to involve a different principle of order selection. Recent crystallographic analysis suggests that its bond ordering need not be accompanied by complete Ir$^{3+}$/Ir$^{4+}$ valence separation \cite{CuIr2S4_emi}. From this perspective, Phase II can be viewed as an orbital-selective bond-ordered state involving molecular-orbital formation rather than as a purely ionic charge-ordered state governed primarily by Coulomb interactions. This interpretation is consistent with the orbitally induced Peierls mechanism originally proposed by Khomskii and Mizokawa \cite{MizoKom}. In this mechanism, the directional $t_{2g}$ orbitals form quasi-one-dimensional electronic states through the shared edges of the IrS$_6$ octahedra, and an instability of these states selects a particular dimerization pattern. The apparent deviation of Phase II from the Anderson condition in a nominal-valence description is therefore not anomalous; it reflects an ordering principle distinct from that associated with the Phase-IV short-bond topology. Phase II and Phase IV should consequently be regarded as distinct classes of bond-ordered state rather than as variants of the same integer-valence charge-ordering pattern.

This distinction provides a possible framework for understanding why the compounds investigated here develop Phase-IV-related monoclinic structures under pressure rather than a Phase-II-type structure. In CuIr$_2$Se$_4$, the greater spatial extent of the Se 4$p$ orbitals relative to the S 3$p$ orbitals may enhance $d$--$p$ hybridization and weaken the quasi-one-dimensional electronic character required for the orbitally induced Peierls mechanism. This would disfavor the Phase-II-type order while allowing a Phase-IV-related structure to become stabilized under pressure. In the Rh-based compounds, the Rh 4$d$ orbitals are less spatially extended than the Ir 5$d$ orbitals. Consistent with this difference, the present DFT calculations yield narrower Rh-derived 4$d$ bands than the corresponding Ir-derived 5$d$ bands. Stronger compression may therefore be required in the Rh-based compounds to generate sufficient orbital overlap for pronounced Rh--Rh bond disproportionation. Pressure also modifies the hybridization between the transition-metal $d$ and chalcogen $p$ states, potentially further altering the balance between the competing ordering tendencies. These considerations provide a plausible rationale for the observed material trends, although direct calculations comparing the relative energies of the Phase-II- and Phase-IV-type structures would be required to establish the microscopic selection mechanism quantitatively.

The material dependence of the structural transition can be discussed on a similar basis. Because the present diffraction data do not determine all atomic positions with sufficient precision to compare metal--metal bond lengths quantitatively across the three compounds, it is more appropriate to base this comparison on the observed lattice response and monoclinic supercell relationship. CuIr$_2$Se$_4$ and CuRh$_2$S$_4$ exhibit closely related anisotropic lattice distortions, while CuRh$_2$Se$_4$ develops a compatible monoclinic supercell despite the pronounced peak broadening. The structural transformation occurs at a lower pressure and over a narrower pressure interval in CuIr$_2$Se$_4$ than in the Rh-based compounds. This trend is consistent with the greater spatial extent of the Ir 5$d$ orbitals, which may allow the orbital overlap associated with short-bond formation to be achieved at a lower pressure. Within the Rh-based pair, replacing S with Se increases the characteristic Rh--Rh separation, consistent with the higher pressure required to stabilize the monoclinic phase in CuRh$_2$Se$_4$. Thus, the observed transition pressures are plausibly governed by the combined effects of the spatial extent of the transition-metal $d$ orbitals and the metal--metal distance scale set by the chalcogen species. The width of the coexistence region alone, however, does not provide a direct quantitative measure of the free-energy difference between the competing phases, because it may also be affected by strain, pressure inhomogeneity, nucleation, and transformation kinetics.

The close correspondence between the structural and transport phase diagrams further indicates that the pressure-induced monoclinic transformation is strongly coupled to the development of insulating-like transport. This coupling is most clearly resolved in CuIr$_2$Se$_4$ and CuRh$_2$S$_4$. In CuIr$_2$Se$_4$, $T_{\mathrm{c}}$ initially increases with pressure but decreases abruptly as the monoclinic phase and insulating-like transport develop. In CuRh$_2$S$_4$, $T_{\mathrm{c}}$ reaches a maximum before the insulating crossover becomes prominent and subsequently decreases more gradually, remaining finite within part of the structural-transition region. These contrasting pressure dependences demonstrate that the competition between superconductivity and the pressure-induced monoclinic state is strongly material dependent. For CuRh$_2$Se$_4$, no well-defined $T^{*}$ can be assigned within the investigated pressure range, and its atomic-scale high-pressure structure remains unresolved. The relationship among bond disproportionation, insulating transport, and superconductivity therefore cannot yet be evaluated on the same basis. Electrical-resistivity and diffraction measurements at higher pressures will be required to clarify this relationship.

Among the three compounds, only CuIr$_2$Se$_4$ exhibits insulating-like transport extending to at least room temperature within the investigated pressure range. This observation indicates that its high-pressure state is associated with a substantially higher characteristic temperature scale than those of the Rh-based compounds. It is consistent with, but does not by itself prove, a larger electronic stabilization of the bond-disproportionated state in CuIr$_2$Se$_4$. Overall, the present results provide a unified framework in which closely related monoclinic structural transformations are strongly coupled to pressure-induced insulating transport, while their transition pressures, characteristic temperature scales, and relationships with superconductivity are controlled by the transition-metal and chalcogen species.

\section{Summary}

In summary, we investigated the pressure-induced structural and electronic evolution of CuIr$_2$Se$_4$, CuRh$_2$S$_4$, and CuRh$_2$Se$_4$ using synchrotron X-ray diffraction and electrical-resistivity measurements under pressure. All three compounds transform from the ambient-pressure cubic spinel phase to closely related monoclinic high-pressure phases. For CuIr$_2$Se$_4$ and CuRh$_2$S$_4$, constrained profile fits using DFT-relaxed structural models support compatibility with Phase-IV-type bond-disproportionated structures. CuRh$_2$Se$_4$ develops a compatible monoclinic supercell, although its detailed short-bond topology remains unresolved because severe preferred orientation, which could not be adequately accounted for by conventional corrections, together with pronounced diffraction-peak broadening and overlap, precluded a reliable refinement of its atomic structure. In the Phase-IV-type structural models, the short-bond topology satisfies the Anderson condition when represented using nominal $B^{3+}/B^{4+}$ labels; these labels serve only as a bookkeeping scheme and do not imply complete integer-valence separation.

The transport properties evolve in close correspondence with these structural transformations. All three compounds exhibit metallic transport at low pressure and develop increasingly insulating-like behavior as the monoclinic phase emerges. The characteristic pressure and temperature scales of this evolution, however, differ markedly among the compounds. In CuIr$_2$Se$_4$, insulating behavior develops abruptly over a relatively narrow pressure range and extends to at least room temperature, whereas the corresponding evolution is more gradual in the Rh-based compounds. We also establish previously unreported bulk superconductivity in CuIr$_2$Se$_4$ at ambient pressure. The zero-resistance transition at 0.29~K is accompanied by an ac diamagnetic response corresponding to an estimated shielding fraction close to 100\%. Superconductivity is progressively suppressed as the pressure-induced monoclinic phase develops, although the manner of this suppression is strongly material dependent.

These results establish a close relationship among the structural transformation, the evolution toward insulating transport, and the suppression of superconductivity in this series. The pronounced material dependence can be understood in terms of the spatial extent of the transition-metal $d$ orbitals and the metal--metal distance and hybridization controlled by the chalcogen species. Elemental substitution thus tunes the stability and pressure evolution of a common Phase-IV-type structural tendency, thereby governing the competition among metallic, superconducting, and pressure-induced insulating states in group-9 spinels.

\begin{acknowledgments}
The authors are grateful to Prof. Takashi Mizokawa for fruitful discussion. The work leading to these results has received funding from the Grant in Aid for Scientific Research (Nos. JP20H02604, JP21K18599, JP22KJ1521, JP23H04104, JP23H04861, JP24H01620, JP24K01329, JP25H01676, JP25K07222, JP26K07021, JP26K00019, JP26K00023, JP26H00599, JP26H00590, JP26K17087) and Japan Science and Technology Agency COI-NEXT Program (No. JPMJPF2221). This work was supported by the Collaborative Research Project of Materials and Structures Laboratory, Institute of Integrated Research, Institute of Science Tokyo. This work was carried out under the Visiting Researcher's Program of the Institute for Solid State Physics, the University of Tokyo (ISSPkyodo-202105-MCBXG-0055, 202111-MCBXG-0002, 202204-MCBXG-0047, 202211-MCBXG-0044 and 202405-MCBXG-0073). Synchrotron XRD experiments were performed at BL02B1 (Proposals No. 2024A1704) and BL02B2 (Proposals No. 2020A1063, No. 2021B1136, No. 2023A1869, No. 2026A1687) and BL10XU (Proposals No. 2020A0530, No. 2021B1120, No. 2022A1166, No. 2022B1126, No. 2023A1116, No. 2023B1114, No. 2024A1151, No. 2024B1733, No. 2025B1172, No. 2025B1675, No. 2026A1116) at SPring-8, Hyogo, Japan. A part of this work was conducted in the University of Electro-Communications Coordinated Center for UEC Research Facilities, supported by “Advanced Research Infrastructure for Materials and Nanotechnology in Japan (ARIM)" of the Ministry of Education, Culture, Sports, Science and Technology (MEXT), under Proposal No. JPMXP1225UE0024.
\end{acknowledgments}

\appendix
\section{Single-crystal X-ray diffraction results}
Table~\ref{tab:table1} summarizes the crystallographic and data-collection parameters for CuRh$_2$S$_4$ and CuRh$_2$Se$_4$ at 100 K; the corresponding diffraction patterns are shown in the insets of Figs.~\ref{fig:Figure3}(a) and~\ref{fig:Figure3}(b). The atomic coordinates and anisotropic atomic displacement parameters of CuRh$_2$S$_4$ are listed in Table~\ref{tab:CuRh2S4_AP} and Table~\ref{tab:adp_CuRh2S4}, respectively, while those of CuRh$_2$Se$_4$ are listed in Table~\ref{tab:CuRh2Se4_AP} and Table~\ref{tab:adp_CuRh2Se4}, respectively.

\begin{table}[!ht]
\caption{\label{tab:table1}
Crystallographic and data-collection parameters of CuRh$_2$S$_4$ and CuRh$_2$Se$_4$ at 100~K.}
\begin{ruledtabular}
\begin{tabular}{lcc}
Parameter & CuRh$_2$S$_4$ & CuRh$_2$Se$_4$ \\
\hline
Temperature (K) 
    & 100 
    & 100 \\
Wavelength (\AA) 
    & 0.31120 
    & 0.31120 \\
Crystal dimensions ($\mu\mathrm{m}$) 
    & $30 \times 10 \times 30$ 
    & $20 \times 10 \times 20$ \\
Space group 
    & $Fd\bar{3}m$ 
    & $Fd\bar{3}m$ \\
$a$ (\AA) 
    & 9.85549(4) 
    & 10.34302(5) \\
$V$ (\AA$^{3}$) 
    & 957.270(6) 
    & 1106.475(9) \\
$Z$ 
    & 8 
    & 8 \\
$F(000)$ 
    & 1464 
    & 2040 \\
$d_{\mathrm{min}}$ (\AA) 
    & 0.28 
    & 0.28 \\
$N_{\mathrm{total,obs}}$ 
    & 49289 
    & 56316 \\
$N_{\mathrm{unique,obs}}$ 
    & 1195 
    & 1353 \\
Average redundancy 
    & 41.2 
    & 41.6 \\
Completeness 
    & 0.994 
    & 0.995 \\
$R_{1}$ 
    & 0.0083 
    & 0.0171 \\
GOF 
    & 0.91 
    & 1.61 \\
\end{tabular}
\end{ruledtabular}
\end{table}

\begin{table}[!ht]
\caption{\label{tab:CuRh2S4_AP}
Atomic coordinates of CuRh$_2$S$_4$ at 100 K and ambient pressure.}
\begin{ruledtabular}
\begin{tabular}{lccccc}
Site & Wyck. & Occ. & $x$ & $y$ & $z$ \\
\hline
Cu & 8$a$  & 1 & $7/8$      & $7/8$      & $7/8$ \\
Rh & 16$d$ & 1 & $1/2$      & $1/2$      & $1/2$ \\
S  & 32$e$ & 1 & 0.259218(6) & $x$    & $x$ \\
\end{tabular}
\end{ruledtabular}
\end{table}

\begin{table}[!ht]
\caption{\label{tab:CuRh2Se4_AP}
Atomic coordinates of CuRh$_2$Se$_4$ at 100 K and ambient pressure.}
\begin{ruledtabular}
\begin{tabular}{lccccc}
Site & Wyck. & Occ. & $x$ & $y$ & $z$ \\
\hline
Cu & 8$a$  & 1 & $7/8$      & $7/8$      & $7/8$ \\
Rh & 16$d$ & 1 & $1/2$      & $1/2$      & $1/2$ \\
Se  & 32$e$ & 1 & 0.259333(5) & $x$    & $x$ \\
\end{tabular}
\end{ruledtabular}
\end{table}

\section{Structural parameters used in the high-pressure structure analyses}
The lattice parameters and structural parameters used for the analyses of CuIr$_2$Se$_4$ and CuRh$_2$S$_4$ are summarized in Table~\ref{tab:table3} and Table~\ref{tab:table4}, respectively. The corresponding refinement results are shown in Fig.~\ref{fig:Figure2}(g) and Fig.~\ref{fig:Figure3}(i). The measurements were performed at 202 K and 11.9 GPa for CuIr$_2$Se$_4$, and at 100 K and 16.0 GPa for CuRh$_2$S$_4$. The lattice parameters were refined, while the fractional atomic coordinates were fixed to the values obtained from the structural optimization calculations. The atomic displacement parameters were fixed to the corresponding ambient-pressure values measured at the respective temperatures. Because the high-pressure structures are highly complex and the diffraction peaks are substantially broadened under pressure, individual reflections cannot be sufficiently resolved, making reliable refinement of the structural parameters difficult.

\section{The temperature dependence of $d\ln\rho/dT$}
\begin{figure}
\includegraphics[width=85mm]{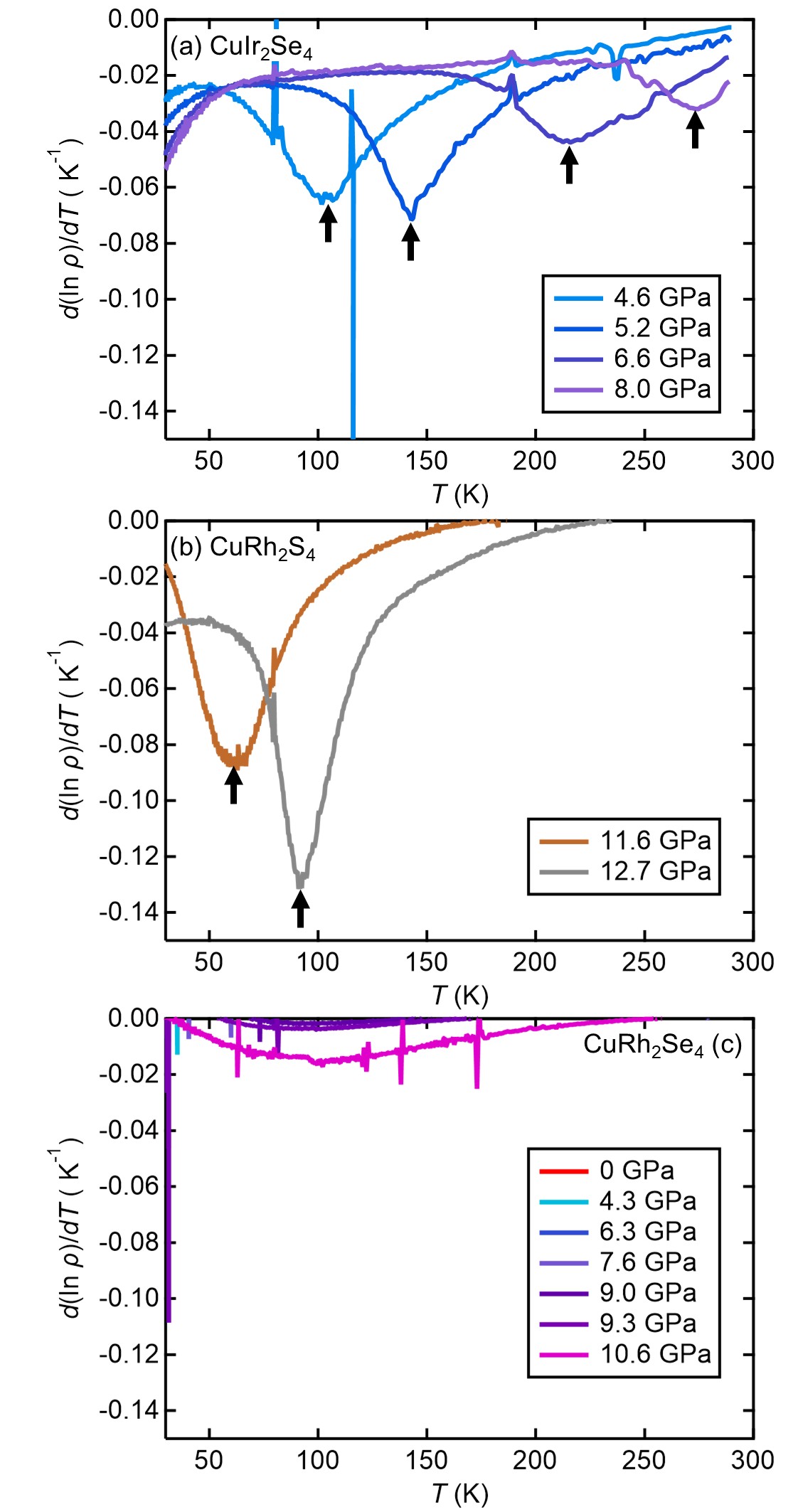}
\caption{\label{fig:Figure8} 
Temperature dependences of $d\ln\rho/dT$ for (a) CuIr$_2$Se$_4$, (b) CuRh$_2$S$_4$, and (c) CuRh$_2$Se$_4$. The arrows in panels (a) and (b) indicate the well-defined minima used to determine $T^{*}$. In panel (c), the curve at 10.6 GPa exhibits a shallow and broad minimum-like feature; however, no $T^{*}$ is assigned because the feature is not sufficiently well defined and the corresponding insulating-like upturn in $\rho$($T$) is weak.}
\end{figure}

Figure~\ref{fig:Figure8} shows the temperature dependences of $d\ln\rho/dT$ calculated from the electrical-resistivity data shown in Figs.~\ref{fig:Figure7}(a)--\ref{fig:Figure7}(c). When a well-defined minimum associated with the insulating crossover is present, we define the characteristic crossover temperature $T^{*}$ as the temperature of this minimum in $d\ln\rho/dT$. In CuRh$_2$Se$_4$, the $d\ln\rho/dT$ curve at 10.6 GPa exhibits a shallow and broad minimum-like feature. Because this feature is much less pronounced than those observed in CuIr$_2$Se$_4$ and CuRh$_2$S$_4$ and is accompanied by only a weak insulating-like upturn in $\rho$($T$), we do not assign a value of $T^{*}$ based on this feature.

\begin{table*}[!ht]
\caption{\label{tab:adp_CuRh2S4}
Anisotropic atomic displacement parameters of CuRh$_2$S$_4$ at 100 K and ambient pressure.}
\begin{ruledtabular}
\begin{tabular}{lcccccc}
Site
& $U_{11}$ (\AA$^2$)
& $U_{22}$ (\AA$^2$)
& $U_{33}$ (\AA$^2$)
& $U_{12}$ (\AA$^2$)
& $U_{13}$ (\AA$^2$)
& $U_{23}$ (\AA$^2$) \\
\hline
Cu
& 0.003638(8)
& 0.003638
& 0.003638
& 0
& 0
& 0 \\
Rh
& 0.002497(6)
& 0.002497
& 0.002497
& $-0.000210(2)$
& $-0.000210$
& $-0.000210$ \\
S
& 0.003201(8)
& 0.003201
& 0.003201
& 0.000046(9)
& 0.000046
& 0.000046 \\
\end{tabular}
\end{ruledtabular}
\end{table*}

\begin{table*}[h]
\caption{\label{tab:adp_CuRh2Se4}
Anisotropic atomic displacement parameters of CuRh$_2$Se$_4$ at 100 K and ambient pressure.}
\begin{ruledtabular}
\begin{tabular}{lcccccc}
Site
& $U_{11}$ (\AA$^2$)
& $U_{22}$ (\AA$^2$)
& $U_{33}$ (\AA$^2$)
& $U_{12}$ (\AA$^2$)
& $U_{13}$ (\AA$^2$)
& $U_{23}$ (\AA$^2$) \\
\hline
Cu
& 0.004695(18)
& 0.004695
& 0.004695
& 0
& 0
& 0 \\
Rh
& 0.003396(12)
& 0.003396
& 0.003396
& $-0.000159(6)$
& $-0.000159$
& $-0.000159$ \\
Se
& 0.003574(12)
& 0.003574
& 0.003574
& 0.000036(7)
& 0.000036
& 0.000036 \\
\end{tabular}
\end{ruledtabular}
\end{table*}

    \begin{table*}
\caption{\label{tab:table3}%
Structural parameters used in the constrained profile fit to the high-pressure diffraction data of CuIr$_2$Se$_4$ at 202 K and 11.9 GPa. The lattice parameters were refined against the diffraction data, whereas the fractional atomic coordinates were obtained from structural-optimization calculations and fixed during the fit. The isotropic atomic displacement parameters were fixed to the corresponding ambient-pressure values. Space group is $C2/c$. Cell parameters are \textit{a} = 24.0603(7) \rm{\AA}, \textit{b} = 7.1416(2) \rm{\AA}, \textit{c} = 11.8023(5) \rm{\AA}, ${\beta}$ =99.590(3)$\tcdegree$. 
The CIF has been deposited with the CCDC (deposition number: 2581642).}
\begin{ruledtabular}
\begin{tabular}{ccccccc}
Site & Wyck. & Occ. & $x$ & $y$ & $z$
& $B_{\mathrm{iso}}$ (\AA$^2$) \\
\hline
Cu1 & 8$f$ & 1 & 0.27617 & 0.27231 & 0.83983 & 0.559 \\
Cu2 & 8$f$ & 1 & 0.03147 & 0.21665 & 0.10233 & 0.559 \\
Ir1 & 8$f$ & 1 & 0.12452 & 0.28932 & 0.61256 & 0.228 \\
Ir2 & 8$f$ & 1 & 0.87563 & 0.47620 & 0.10943 & 0.228 \\
Ir3 & 8$f$ & 1 & 0.87668 & 0.02211 & 0.62564 & 0.228 \\
Ir4 & 4$d$ & 1 & 1/4 & 1/4 & 1/2 & 0.228 \\
Ir5 & 4$e$ & 1 & 0 & 0.71361 & 1/4 & 0.228 \\
Se1 & 8$f$ & 1 & 0.43633 & 0.23145 & 0.06696 & 0.276 \\
Se2 & 8$f$ & 1 & 0.18277 & 0.23218 & 0.32259 & 0.276 \\
Se3 & 8$f$ & 1 & 0.43596 & 0.44179 & 0.31029 & 0.276 \\
Se4 & 8$f$ & 1 & 0.43395 & 0.01544 & 0.80348 & 0.276 \\
Se5 & 8$f$ & 1 & 0.69547 & 0.00265 & 0.56177 & 0.276 \\
Se6 & 8$f$ & 1 & 0.68324 & 0.47076 & 0.05983 & 0.276 \\
Se7 & 8$f$ & 1 & 0.56433 & 0.27862 & 0.93012 & 0.276 \\
Se8 & 8$f$ & 1 & 0.82196 & 0.27698 & 0.69090 & 0.276 \\

\end{tabular}
\end{ruledtabular}
\end{table*}

    \begin{table*}
\caption{\label{tab:table4}%
Structural parameters used in the constrained profile fit to the high-pressure diffraction data of CuRh$_2$S$_4$ at 100 K and 16.0 GPa. The lattice parameters were refined against the diffraction data, whereas the fractional atomic coordinates were obtained from structural-optimization calculations and fixed during the fit. The isotropic atomic displacement parameters were fixed to the corresponding ambient-pressure values. Space group is $C2/c$. Cell parameters are \textit{a} = 22.6431(11) \rm{\AA}, \textit{b} = 6.7346(3) \rm{\AA}, \textit{c} = 11.1920(10) \rm{\AA}, ${\beta}$ =99.660(5)$\tcdegree$. The CIF has been deposited with the CCDC (deposition number: 2581677).}
\begin{ruledtabular}
\begin{tabular}{ccccccc}
Site & Wyck. & Occ. & $x$ & $y$ & $z$
& $B_{\mathrm{iso}}$ (\AA$^2$) \\
\hline
Cu1 & 8$f$ & 1 & 0.27772 & 0.26574 & 0.84096 & 0.173 \\
Cu2 & 8$f$ & 1 & 0.03110 & 0.22395 & 0.10025 & 0.173 \\
Rh1 & 8$f$ & 1 & 0.12469 & 0.28059 & 0.61548 & 0.092 \\
Rh2 & 8$f$ & 1 & 0.87547 & 0.48046 & 0.11299 & 0.092 \\
Rh3 & 8$f$ & 1 & 0.87667 & 0.01790 & 0.62522 & 0.092 \\
Rh4 & 4$d$ & 1 & 1/4 & 1/4 & 1/2 & 0.092 \\
Rh5 & 4$e$ & 1 & 0 & 0.71938 & 1/4 & 0.092 \\
S1  & 8$f$ & 1 & 0.43635 & 0.23444 & 0.06843 & 0.204 \\
S2  & 8$f$ & 1 & 0.18357 & 0.23421 & 0.32271 & 0.204 \\
S3  & 8$f$ & 1 & 0.43594 & 0.45131 & 0.30905 & 0.204 \\
S4  & 8$f$ & 1 & 0.43393 & 0.00997 & 0.80468 & 0.204 \\
S5  & 8$f$ & 1 & 0.19338 & 0.49900 & 0.56071 & 0.204 \\
S6  & 8$f$ & 1 & 0.68383 & 0.47518 & 0.05887 & 0.204 \\
S7  & 8$f$ & 1 & 0.56502 & 0.27338 & 0.93232 & 0.204 \\
S8  & 8$f$ & 1 & 0.82077 & 0.27141 & 0.69089 & 0.204 \\

\end{tabular}
\end{ruledtabular}
\end{table*}

\clearpage

\nocite{*}

\bibliography{references}
\end{document}